\documentclass[aip,preprint,pop]{revtex4-2}
\usepackage{graphicx}
\usepackage{amsmath,amssymb,amsfonts,bm}
\usepackage{caption}
\usepackage[subrefformat=parens]{subcaption}
\newcommand{\bv}{\boldsymbol}
\newcommand{\pd}{\partial}
\newcommand{\td}{{\rm d}}
\newcommand{\nab}{\nabla}
\newcommand{\pr}{\prime}
\newcommand{\dd}{\cdot\nab}
\newcommand{\rot}{\nabla\times}
\newcommand{\rmi}{{\rm i}\,}
\newcommand{\rme}{{\rm e}}
\newcommand{\vtheta}{\vartheta}
\newcommand{\vh}{{\bm h}}
\newcommand{\hvtheta}{\hat{\bv\theta}}
\newcommand{\vphi}{\varphi}
\newcommand{\muo}{\mu_{0}}
\newcommand{\vB}{{\bm B}}
\newcommand{\vu}{{\bm u}}
\newcommand{\vk}{{\bm k}}
\renewcommand{\vr}{{\bm r}}
\newcommand{\hrvz}{\hat{\bm z}}
\newcommand{\vrme}{{\bf e}}
\newcommand{\dis}{\displaystyle}

\begin{document}

%\preprint{}

\title{
Hamiltonian structure of single-helicity, incompressible magnetohydrodynamics
and application to magnetorotational instability
}

\author{M. Furukawa}
\email[]{furukawa@tottori-u.ac.jp}
%\thanks{}
\affiliation{
Faculty of Engineering, Tottori Univ., 
Minami 4-101, Koyama-cho, Tottori-shi, Tottori 680-8552, Japan
}

\author{M. Hirota}
%\thanks{}
\affiliation{
Institute of Fluid Science, Tohoku Univ., 2-1-1 Katahira, Aoba-ku, Sendai, Miyagi 980-8577, Japan
}

\date{\today}

\begin{abstract}
A four-field reduced model of single helicity, incompressible MHD is
 derived in cylindrical geometry.  
An appropriate set of noncanonical variables is found, and
 the Hamiltonian, the Lie--Poisson bracket and the Casimir invariants
 are clarified.  Detailed 
 proofs of properties of the Lie--Poisson bracket, (i) antisymmetry,
 (ii) Leibniz 
 rule, and (iii) Jacobi identity, are given.  Two applications are
 presented: the first is that the local 
 dispersion relation of  axisymmetric magnetohydrodynamics (MRI) is
 properly reproduced, and the second is that linear stability analyses
 including negative-energy MRI were successfully performed.
\end{abstract}


\maketitle

%%%%%%%%%%%%%%%%%%%%%%%%%%%%
%%%%%%%%%%%%%%%%%%%%%%%%%%%%
%%%%%%%%%%%%%%%%%%%%%%%%%%%%

\section{Introduction}
\label{sec:introduction}

Ideal fluid and magnetohydrodynamic (MHD) systems have 
noncanonical Hamiltonian structure.
The dynamics is described by a Hamiltonian and a Poisson bracket.
Because the dynamical variables are noncanonical, the Poisson bracket
has a null space that corresponds to Casimir invariants.
Many aspects of the system such as equilibrium and stability can be
understood in the context of dynamical systems. 

The Hamiltonian structure of the ideal fluid and MHD in
three-dimensional space was found in 
Ref.~\onlinecite{Morrison-1980}.  
Also for reduced MHD systems, 
\cite{Strauss-1976, Strauss-1977}
the Hamiltonian structure was clarified in
Ref.~\onlinecite{Morrison-Hazeltine-1984}.   
Reviews of the Hamiltonian description of a variety of systems can be
found, for example, in Refs.~\onlinecite{Holm-1985, Morrison-1998}.

Such reduced models have played essential roles 
in studies of magnetically confined fusion plasmas. 
The reduction relies basically on an assumption that the aspect ratio is
large or the wave length parallel to magnetic field is 
much larger than that perpendicular to the magnetic field.
The reduced MHD models are often valid under strong ambient magnetic field.

An accretion disk, however, can be a different situation, where 
a magnetorotational instability (MRI) can occur under presence of 
a weak ambient magnetic field 
\cite{Velikhov-1959, Chandrasekhar-1960, B-H-1991, B-H-1998}.
The simplest situation for the MRI to occur is 
that a homogeneous magnetic field is in the $z$ direction of the
cylindrical coordinates $(r, \theta, z)$, and the plasma rotates in the 
$\theta$ direction.  Naively, this situation may be realized by using
the low-beta reduced MHD in cylindrical geometry, leaving the
assumption of strong ambient magnetic field  made in the derivation
aside.  
However, MRI cannot be expressed by the low-beta reduced
MHD model because it only retains the shear Alfv\'en wave.
MRI needs a coupling between the shear Alfv\'en wave and the slow
magnetosonic wave. 

In the present paper, we derive 
a four-field reduced MHD model that is a minimum model to MRI,
and elucidate its Hamiltonian structure.
The model is derived by taking appropriate components of 
the ideal MHD equations under assumptions of 
single helicity dynamics and incompressibility.  
No assumption is made on the aspect ratio and plasma beta.
A choice of noncanonical variables is essential in clarifying 
the Hamiltonian structure. 

The rest of this paper is organized as follows. 
The governing equations of the four-field reduced MHD model 
under assumptions of single helicity and incompressibility 
are derived in Sec.~\ref{sec:derivationGoverningEquations}.
Then, an appropriate set of noncanonical variables, 
together with the Hamiltonian, the Lie--Poisson bracket, and the Casimir
invariants are found in 
Sec.~\ref{sec:HamiltonianStructure}.
Moreover, important properties of the Lie--Poisson bracket,
(i) antisymmetry, (ii) Leibniz rule, and (iii) Jacobi identity,
are proven.
In Sec.~\ref{sec:application}, two applications are presented.  
The first one demonstrates that the derived model can successfully
reproduce the well-known local dispersion relation of axisymmetric MRI.
The other application shows linear stability analysis of MRI including
negative energy modes, which was also studied in 
Refs.~\onlinecite{Khalzov-2006, Khalzov-2008, Ilgisonis-2009} 
by using the Frieman--Rotenberg equation \cite{F-R-1960}.
Conclusions are given in Sec.~\ref{sec:conclusion}.
In Appendix~\ref{sec:energyCalculation}, 
a formula for calculating perturbed energy of MRI is given.

\section{Derivation of governing equations}
\label{sec:derivationGoverningEquations}

Let us consider a cylindrical plasma with a minor radius $a$ and 
an axial length $2 \pi R_{0}$.
If we consider single-helicity dynamics, a physical quantity $f$ can be
expressed in a Fourier series as 
\begin{equation}
 f(r, \theta, z, t) 
= 
\sum_{\ell = -\infty}^{\infty} f_{\ell}(r, t) 
\rme^{\rmi \ell ( M \theta + N \zeta )}
\label{eq:FourierExpansion}
\end{equation}
in cylindrical coordinates $(r, \theta, z)$.
Here, $\rmi$ is the imaginary unit, 
$\zeta := z / R_{0}$ is an angle coordinate in the axial direction, 
$\ell$ is an integer, 
$M$ and $N$ are principal poloidal and toroidal mode numbers,
respectively. 
We assume $N \neq 0$.

It is noted that the ``single-helicity'' means that the
dynamics includes a single family of Fourier modes expressed by 
$\ell M$ and $\ell N$.  
It is not the usual helicity that is defined by $\vu \cdot \rot \vu$ for
a velocity field $\vu$.

By introducing a set of coordinates $(r, \vtheta, \alpha)$ via
\begin{align}
 \vtheta
&:= 
 \theta,
\label{eq:vtheta}
\\
 \alpha 
&:=
 R_{0} \left( \frac{M}{N} \theta + \zeta \right)
\quad
= 
 \frac{M}{K} \theta + z
\label{eq:alpha}
\end{align}
with $K := N / R_{0}$,
Eq.~(\ref{eq:FourierExpansion}) can be expressed as 
\begin{equation}
 f(r, \alpha, t) 
= 
\sum_{\ell = -\infty}^{\infty} f_{\ell}(r, t) 
\rme^{\rmi \ell K \alpha}.
\label{eq:FourierExpansion2}
\end{equation}
Thus any physical quantity does not depend on $\vtheta$
in the single-helicity dynamics.

Now, let us introduce an incompressible vector field 
\begin{equation}
 \vh 
:=
 \frac{1}{K_{0}^{2} r^{2}} ( -K r \hvtheta + M \hrvz )
\label{eq:vh}
\end{equation}
with
\begin{equation}
 K_{0}^{2} r^{2} 
:= 
 M^{2} + K^{2} r^{2}
=
 \frac{1}{ | \vh |^{2} },
\label{eq:K02r2}
\end{equation}
where $\hvtheta$ and $\hrvz$ are unit vectors in $\theta$ and $z$
directions, respectively.
Then, for an arbitrary function $f(r, \alpha , t)$, we have
\begin{equation}
 \vh \dd f = 0.
\label{eq:hddf0}
\end{equation}

Using the vector field $\vh$, an incompressible flow velocity field and
a magnetic field can be expressed as 
\begin{align}
 \vu 
&=
 \vh \times \nab \vphi(r, \alpha, t) + u_{h}(r, \alpha, t) \vh,
\label{eq:vu}
\\
 \vB 
&= 
 \nab \psi(r, \alpha, t) \times \vh + B_{h}(r, \alpha, t) \vh.
\label{eq:vB}
\end{align}

As usual in reduced MHD models, let us introduce a Lie bracket
as 
\begin{equation}
 [f, g]
:= 
 \vh \cdot \nab f \times \nab g
\label{eq:PoissonBracket}
\end{equation}
for arbitrary functions $f(r, \alpha, t)$ and $g(r, \alpha, t)$.
The Lie bracket is shown to satisfy the following three properties:
\begin{enumerate}
 \item [(i)] Antisymmetry:
\begin{equation}
 [f, g]
=
 -[g, f],
\label{eq:PoissonBracketAntisymmetry}
\end{equation}
 \item [(ii)] Leibniz rule:
\begin{align}
 [f g, h]
&=
 g [f, h] + f [g, h],
\label{eq:PoissonBracketLeibnitzRule1}
\\
 [f, g h]
&=
 g [f, h] + h [f, g],
\label{eq:PoissonBracketLeibnitzRule2}
\end{align}
 \item [(iii)] Jacobi identity:
\begin{equation}
 [f, [g, h]] + [g, [h, f]] + [h, [f, g]] = 0.
\label{eq:PoissonBracketJacobiIdentity}
\end{equation}
\end{enumerate}
Here $h(r, \alpha, t)$ is another arbitrary function.
Furthermore, the Lie bracket satisfies the following property
\begin{equation}
 \int \td V \, f [g, h]
=
 \int \td V \,  h [f, g]
%\label{eq:PoissonBracketProperty1}
=
 \int \td V \,  g [h, f]
%\label{eq:PoissonBracketProperty2}
\label{eq:PoissonBracketProperty}
\end{equation}
under an appropriate boundary condition, 
where the volume integral is taken inside the plasma.
The appropriate boundary condition means that any boundary term that
arise from an integration by parts vanishes.

By using the Poisson bracket, the following relations are obtained
for an arbitrary physical quantity $f$:
\begin{align}
 \vu \dd f
&= 
 [\vphi, f],
\label{eq:uddf}
\\
 \vB \dd f
&= 
 -[\psi, f].
\label{eq:Bddf}
\end{align}

The basic equations for deriving the reduced model are ideal MHD
equations, which are 
normalized by using a typical length $L_{0}$,
a typical magnetic field $B_{0}$, a typical mass density $\rho_{0}$,
a typical velocity $v_{0} := B_{0} / \sqrt{\muo \rho_{0}}$ 
with $\muo$ being the vacuum permeability, and a typical time 
$t_{0} := L_{0} / v_{0}$.
Explicitly, they are the equation of motion
\begin{equation}
 \left(
   \frac{\pd }{\pd t} + \vu \dd
 \right)  \vu
=
 ( \rot \vB ) \times \vB - \nab p
\label{eq:eqMotion}
\end{equation}
and its rotation, the vorticity equation
\begin{equation}
  \frac{\pd ( \rot \vu )}{\pd t} 
=
 \rot ( \vu \times ( \rot \vu ) )
 + \rot ( ( \rot \vB ) \times \vB ),
\label{eq:vorticityEq}
\end{equation}
the ideal Ohm's law 
\begin{equation}
 \frac{\pd \bv{A}}{\pd t}
=
  \vu \times \vB - \nab \phi 
\label{eq:OhmLaw}
\end{equation}
and its rotation, the induction equation
\begin{equation}
 \frac{\pd \vB}{\pd t}
=
 \rot ( \vu \times \vB ).
\label{eq:inductionEq}
\end{equation}
Here, 
$\vu$ is the fluid velocity, $\vB$ is the magnetic field,
$p$ is the pressure, 
$\bv{A}$ is the vector potential, and 
$\phi$ is the electrostatic potential.
The mass density is constant due to incompressibility, 
and the normalized mass density is $1$.

Taking a dot product of the vorticity equation (\ref{eq:vorticityEq})
with $\vh$, we obtain
\begin{align}
&
 \frac{\pd}{\pd t} 
 \left(
   \mathcal{L} \vphi
- \frac{2 M K}{( K_{0}^{2} r^{2} )^{2}} u_{h}
 \right)
\nonumber
\\
&= 
- \left[ u_{h},  \frac{1}{K_{0}^{2} r^{2}} u_{h} 
  \right] 
- \left[ 
   \vphi,  
   \mathcal{L} \vphi
  \right]
+ \left[
   \vphi,
   \frac{2 M K}{( K_{0}^{2} r^{2} )^{2}} u_{h} 
  \right]
- \frac{2 M K}{( K_{0}^{2} r^{2} )^{2}}
  [ u_{h}, \vphi ]
\nonumber
\\
& \quad
+ \left[ 
   B_{h},  
   \frac{1}{K_{0}^{2} r^{2}} B_{h} 
  \right] 
+ \frac{1}{\muo \rho}
  \left[ 
   \psi,  
   \mathcal{L} \psi
  \right]
%\nonumber
%\\
%& \quad
+ \left[
    \psi, 
    \frac{2 M K}{( K_{0}^{2} r^{2} )^{2}} B_{h} 
  \right]
- \frac{2 M K}{( K_{0}^{2} r^{2} )^{2}}
  [ B_{h}, \psi ],
\label{eq:h.vorticityEq0}
\end{align}
where
\begin{equation}
 \mathcal{L}
:= 
 \frac{1}{K r} 
   \frac{\pd}{\pd r} 
    \left( \frac{K r}{K_{0}^{2} r^{2}} \frac{\pd }{\pd r} \right)
+ \frac{1}{K^{2} r^{2}} 
    \frac{\pd^{2} }{\pd \alpha^{2}}.
\label{eq:calL-definition}
\end{equation}

Next, a dot product of the equation of motion (\ref{eq:eqMotion})
with $\vh$ yields
\begin{equation}
 \frac{\pd u_{h}}{\pd t}
=
 [ u_{h}, \vphi] 
 + [ B_{h}, \psi ].
 \label{eq:h.eqMotion}
\end{equation}
Note that $\vh \dd p(r, \alpha, t) \equiv 0$.
Multiplying (\ref{eq:h.eqMotion}) with 
$2 M K / ( K_{0}^{2} r^{2} )^{2}$
and adding it to Eq.~(\ref{eq:h.vorticityEq0}),
we obtain 
\begin{align}
&
 \frac{\pd}{\pd t} 
 \left(
   \mathcal{L} \vphi
 \right)
\nonumber
\\
&= 
- \left[ u_{h},  \frac{1}{K_{0}^{2} r^{2}} u_{h} 
  \right] 
- \left[ 
   \vphi,  
   \mathcal{L} \vphi
  \right]
+ \left[
   \vphi,
   \frac{2 M K}{( K_{0}^{2} r^{2} )^{2}} u_{h} 
  \right]
\nonumber
\\
& \quad
+ \left[ 
   B_{h},  
   \frac{1}{K_{0}^{2} r^{2}} B_{h} 
  \right] 
+ \left[ 
   \psi,  
   \mathcal{L} \psi
  \right]
+ \left[
    \psi, 
    \frac{2 M K}{( K_{0}^{2} r^{2} )^{2}} B_{h} 
  \right].
 \label{eq:h.vorticityEq}
\end{align}

Third, let us take a dot product of the ideal Ohm's law (\ref{eq:OhmLaw})
with $\vh$.
The components of the vector potential $\bv{A}$ and the magnetic field $\vB$
can be related as follows.
The magnetic field is expressed as 
\begin{equation}
 \vB 
=
  \frac{1}{K r} \frac{\pd \psi}{\pd \alpha} 
 \vrme_{r}
- \frac{1}{K_{0}^{2} r^{2}} 
 \frac{1}{r}
 \left(
   M \frac{\pd \psi}{\pd r}
  + K r B_{h}
 \right) 
 \vrme_{\vtheta}
- \frac{1}{K r} \frac{\pd \psi}{\pd r}
 \vrme_{\alpha},
\label{eq:BComponents}
\end{equation}
where $\vrme_{r} := \pd \vr / \pd r$,
$\vrme_{\vtheta} := \pd \vr / \pd \vtheta$,
and $\vrme_{\alpha} := \pd \vr / \pd \alpha$
with $\vr$ being a position vector.
On the other hand, 
if we express 
the vector potential as
$\bv{A} = A_{r} \nab r + A_{\vtheta} \nab \vtheta + A_{\alpha} \nab \alpha$,
where where $A_{r} = \bv{A} \cdot \vrme_{r}$,
$A_{\vtheta} = \bv{A} \cdot \vrme_{\vtheta}$,
and $A_{\alpha} = \bv{A} \cdot \vrme_{\alpha}$
are covariant components, 
$\rot \bv{A}$
reads 
\begin{equation}
 \rot \bv{A}
=
 -\frac{1}{r}
  \frac{\pd A_{\vtheta}}{\pd \alpha}
  \vrme_{r}
+
 \frac{1}{r}
 \left(
   \frac{\pd A_{r} }{\pd \alpha}
 - \frac{\pd A_{\alpha} }{\pd r}
 \right)
 \vrme_{\vtheta}
+
 \frac{1}{r}
   \frac{\pd A_{\vtheta} }{\pd r}
 \vrme_{\alpha}.
\label{eq:AComponents}
\end{equation}
By comparing Eqs.~(\ref{eq:BComponents})
and (\ref{eq:AComponents}),
the components can be related as 
\begin{align}
 A_{r}
&=
 -\frac{K}{K_{0}^{2} r^{2}} \int B_{h} \td \alpha,
\label{eq:Ar}
\\
 A_{\vtheta}
&=
 -\frac{1}{K} \psi,
\label{eq:Avtheta}
\\
 A_{\alpha}
&=
 \int \frac{M}{K_{0}^{2} r^{2}} \frac{\pd \psi}{\pd r} \td r.
\label{eq:Aalpha}
\end{align}
Then,
$\vh \cdot \bv{A} = \psi / ( K_{0}^{2} r^{2} )$.
The resultant equation is obtained as 
\begin{equation}
 \frac{\pd \psi}{\pd t}
=
 [ \psi, \vphi ].
 \label{eq:h.OhmLaw}
\end{equation}
Note that $\vh \dd \phi(r, \alpha, t) \equiv 0$.

Last equation is obtained by a dot product of the induction equation
(\ref{eq:inductionEq}) with $\vh$ as
\begin{equation}
 \frac{\pd B_{h}}{\pd t}
=
 - K_{0}^{2} r^{2}
   \left[
   \psi, 
   \frac{1}{K_{0}^{2} r^{2}} u_{h}
  \right]
 - K_{0}^{2} r^{2}
   \left[
    \vphi,
    \frac{1}{K_{0}^{2} r^{2}} B_{h}
   \right]
 + \frac{2 M K}{K_{0}^{2} r^{2}} [ \vphi, \psi ].
 \label{eq:h.inductionEq}
\end{equation}

The energy of the system
\begin{align}
  E 
&=
 \frac{1}{2}
 \int \td V \,
 \left(  
    | \vu |^{2} + | \vB |^{2}
 \right)
\nonumber
\\
&=
 \frac{1}{2} 
 \int \td V \,
 \frac{1}{K_{0}^{2} r^{2}} 
 \left(
 | \nab \vphi |^{2} + u_{h}^{2}
+ | \nab \psi |^{2} + B_{h}^{2}
 \right),
\label{eq:energy}
\end{align}
is conserved according to Eqs.~(\ref{eq:h.vorticityEq}), 
(\ref{eq:h.eqMotion}), (\ref{eq:h.OhmLaw}), and (\ref{eq:h.inductionEq}).
%Here, the volume integral is taken inside the plasma volume.

\section{Hamiltonian structure}
\label{sec:HamiltonianStructure}

\subsection{Hamiltonian and Lie--Poisson bracket}
\label{subsec:HamiltonianLiePoissonBracket}

The single-helicity, incompressible MHD model derived in
Sec.~\ref{sec:derivationGoverningEquations} can be written in the
Hamiltonian form with noncanonical variables.
Let us introduce a set of phase-space variables
$v = ( v^{1}, v^{2}, v^{3}, v^{4} )^{\mathrm{T}}
:= (U, u_{h}, \psi, B_{h}^{\star})^{\mathrm{T}}$.
Here, $U$ and $B_{h}^{\star}$ are defined as follows:
\begin{align}
 U 
&:=
 \mathcal{L} \vphi,
\label{eq:vorticity}
\\
 B_{h}^{\star}
&:=
 \frac{1}{K_{0}^{2} r^{2}}
 \left(
   B_{h} + \frac{2 M K}{K_{0}^{2} r^{2}} \psi
 \right).
\label{eq:Bhst}
\end{align}

The Hamiltonian is taken to be the energy inside the plasma as
\begin{equation}
 H[v]
=
 \frac{1}{2}
 \int \td V \,
 \left(
  - U ( \mathcal{L}^{-1} U )
  + g u_{h}^{2} 
  - \psi ( \mathcal{L} \psi )
%  + K_{0}^{2} r^{2} \left(
  + \frac{1}{g} \left(
     B_{h}^{\star} + f \psi
                    \right)^{2}
 \right),
 \label{eq:Hamiltonian}
\end{equation}
where 
\begin{align}
 f(r) 
:= & \,
 \vh \cdot \rot \vh
\label{eq:f-definition-1}
\\
= & \,
 -\frac{2 M K}{( K_{0}^{2} r^{2} )^{2}},
\label{eq:f-definition-2}
\\
 g(r)
:= & \,
 | \vh |^{2}
\label{eq:g-definition-1}
\\
= & \,
 \frac{1}{K_{0}^{2} r^{2}}
\label{eq:g-definition-2}
\end{align}
are introduced for simplicity of notation.
Then, functional derivatives of $H[v]$ are given by
\begin{align}
 \frac{\delta H}{\delta U}
&=
 - \vphi,
 \label{eq:dHdU}
\\
 \frac{\delta H}{\delta u_{h}}
&=
 g u_{h},
 \label{eq:dHduh}
\\
 \frac{\delta H}{\delta \psi}
&=
 - \mathcal{L} \psi 
 + f B_{h},
 \label{eq:dHdpsi}
\\
 \frac{\delta H}{\delta B_{h}^{\star}}
&=
 B_{h}.
 \label{eq:dHdBhst}
\end{align}

The Lie--Poisson bracket 
for arbitrary functionals $F[v]$ and $G[v]$
is defined by
\begin{align}
 \{ F , G \}
&=
 \left\langle
     ( U + f u_{h} ) ,
     \left[ 
        \frac{\delta F}{\delta U} ,
        \frac{\delta G}{\delta U}
     \right]
 \right\rangle
+\left\langle
   u_{h}  ,
   \left(
      \left[ 
         \frac{\delta F}{\delta U} ,
         \frac{\delta G}{\delta u_{h}}
      \right]
     +\left[
          \frac{\delta F}{\delta u_{h}} ,
          \frac{\delta G}{\delta U}
      \right]
   \right)
 \right\rangle
\nonumber
\\
& \hspace{5mm}
+\left\langle
   \psi  ,
   \left(
      \left[
          \frac{\delta F}{\delta U} ,
          \frac{\delta G}{\delta \psi}
      \right]
     +\left[
          \frac{\delta F}{\delta \psi} ,
          \frac{\delta G}{\delta U}
      \right]
     +\left[
         \frac{\delta F}{\delta u_{h}} ,
         \frac{\delta G}{\delta B_{h}^{\star}}
      \right]
     +\left[
         \frac{\delta F}{\delta B_{h}^{\star}} ,
         \frac{\delta G}{\delta u_{h}}
      \right]
   \right)
 \right\rangle
\nonumber
\\
& \hspace{5mm}
+\left\langle
   g B_{h}  ,
   \left(
      \left[
          \frac{\delta F}{\delta U} ,
          \frac{\delta G}{\delta B_{h}^{\star}}
      \right]
     +\left[
          \frac{\delta F}{\delta B_{h}^{\star}} ,
          \frac{\delta G}{\delta U}
      \right]
   \right)
 \right\rangle
 \label{eq:LiePoissonBracket1}
\\
&=
 \int \td V \,
 \frac{\delta F}{\delta v^{i}} 
 \mathcal{J}^{ij}
 \frac{\delta G}{\delta v^{j}},
 \label{eq:LiePoissonBracket2}
\end{align}
where the angle bracket denotes an inner product, and 
the Poisson tensor is given by 
\begin{align}
 \mathcal{J}
= & \,
 ( \mathcal{J}^{ij} )
\nonumber
\\
:= & \,
 \begin{pmatrix}
   \left[ \circ , U + f u_{h}  \right]
   &
   \left[ \circ , u_{h} \right]
   &
   \left[ \circ , \psi  \right]
   &
   \left[ \circ , g B_{h}  \right]
   \\
   \left[ \circ , u_{h} \right]
   &
   0
   &
   0
   &
   \left[ \circ , \psi  \right]
   \\
   \left[ \circ , \psi  \right]
   &
   0
   &
   0
   &
   0
   \\
   \left[ \circ , g B_{h}  \right]
   &
   \left[ \circ , \psi  \right]
   &
   0
   & 
   0  
 \end{pmatrix}.
 \label{eq:PoissonTensor}
\end{align}
Then the evolution equation for $v$ is written as 
\begin{align}
 \frac{\pd v^{i}}{\pd t}  
&=
 \mathcal{J}^{ij} \frac{\delta H}{\delta v^{j}}
 \label{eq:evolutionEq1}
\\
&=
 \{ v^{i} , H  \}.
 \label{eq:evolutionEq2}
\end{align}

The single-helicity dynamics focused in this paper forms an invariant
manifold in the phase space of the dynamics of the full set of ideal MHD,
for which the Hamiltonian structure was derived in
Ref.~\onlinecite{Morrison-1980}.  However, it is not trivial what
phase-space variables are appropriate for a reduced model.
A naive choice cannot lead to a Lie--Poison bracket.  
The phase-space variables $v$ 
in the present model 
includes the magnetic flux function $\psi$, of which spatial derivatives
correspond to the components of the magnetic field.  
The components of the magnetic field are part of the phase-space variables 
of the full set of ideal MHD.
Therefore, a simple variable transformation
cannot drive the Lie--Poisson bracket of the present four-field model.  

The Lie--Poisson bracket can be shown to satisfy the following three 
properties:
\begin{enumerate}
 \item [(i)] Antisymmetry:
\begin{equation}
 \{ F, G \} 
=
 -\{ G, F \},
\label{eq:LiePoissonBracketAntisymmetry}
\end{equation}
 \item [(ii)] Leibniz rule:
\begin{align}
 \{ F G , H \}
&=
G \{ F, H \}
+ F \{ G, H \},
\label{eq:LiePoissonBracketLeibnizRule1}
\\
 \{ F , G H \}
&=
G \{ F, H \}
+ H \{ F, G \},
\label{eq:LiePoissonBracketLeibnizRule2}
\end{align}
 \item [(iii)] Jacobi identity:
\begin{equation}
  \{ F, \{ G, H \} \}
+ \{ G, \{ H, F \} \}
+ \{ H, \{ F, G \} \}
= 0,
\label{eq:LiePoissonBracketJacobiIdentity}
\end{equation}
\end{enumerate}
where $H[v]$ is another arbitrary functional.
Note that $H[v]$ here is not the Hamiltonian 
(\ref{eq:Hamiltonian}).
The detailed proofs are given in 
the next subsection.

Note that $f(r)$, defined in Eq.~(\ref{eq:f-definition-1}), 
is helicity of $\vh$, which is different from the one in the
``single-helicity''.  
This vanishes when the system is axisymmetric with $M = 0$ or 
is translationally symmetric in $z$ with $K = 0$
as seen in Eq.~(\ref{eq:f-definition-2}).
As we have seen, and will observe in the proofs of properties of the
Lie--Poisson bracket, nonzero $f$ remains in the helically symmetric
cases, for which it is not trivial to show the Hamiltonian structure
especially the Jacobi identity.

\subsection{Proof of properties of Lie--Poisson bracket}
\label{subsec:LiePoissonBracketProperties}

\subsubsection{Antisymmetry}
\label{subsubsec:ProofLiePoissonBracketAntisymmetry}

Let us write the Poisson tensor 
$\mathcal{J} = ( \mathcal{J}^{ij} ) = ( [ \circ, J^{ij} ] )$
with $J^{ij} = J^{ji}$.
Then, the antisymmetry can be shown directly as 
\begin{align}
 \{ F, G \}
&=
 \int \td V \,
 \frac{\delta F}{\delta v^{i}} 
 \left[ \frac{\delta G}{\delta v^{j}}, J^{ij} \right] 
\nonumber
\\
&=
-\int \td V \,
 \frac{\delta G}{\delta v^{j}} 
 \left[ \frac{\delta F}{\delta v^{i}}, J^{ij} \right] 
\nonumber
\\
&=
-\int \td V \,
 \frac{\delta G}{\delta v^{i}} 
 \left[ \frac{\delta F}{\delta v^{j}}, J^{ji} \right] 
\nonumber
\\
&=
-\{ G, F \}.
\label{}
\end{align}
Here, Eq.~(\ref{eq:PoissonBracketProperty}) and 
the antisymmetry (\ref{eq:PoissonBracketAntisymmetry}) 
are used in the second equality,
and the dummy indices $i$ and $j$ are interchanged 
in the third equality.
In the last equality $J^{ij} = J^{ji}$ is used.

\subsubsection{Leibniz rule}
\label{subsubsec:ProofLiePoissonBracketLeibnizRule}

According to a property of the Fr\'{e}chet derivative
\begin{equation}
 \frac{ \delta{ (FG)} }{\delta v^{i}}
=
 \frac{\delta F}{\delta v^{i}} G
 +  F \frac{\delta G}{\delta v^{i}},
\end{equation}
it is easily shown that 
\begin{align}
 \left\{
   FG, H
 \right\}
&=
 \int \td V \,
   \frac{ \delta{ (FG)} }{\delta v^{i}}
   \mathcal{J}^{ij}
   \frac{\delta H}{\delta v^{j}}
%\nonumber
%\\
%&=
% \int \td V \,
%   \left(
%      \frac{\delta F}{\delta v^{i}} G
%      +  F \frac{\delta G}{\delta v^{i}}
%   \right)
%   \mathcal{J}^{ij}
%   \frac{\delta H}{\delta v^{j}}
\nonumber
\\
&=
 G
 \int \td V \,
   \frac{\delta F}{\delta v^{i}}
   \mathcal{J}^{ij}
   \frac{\delta H}{\delta v^{j}}
+
 F
 \int \td V \,
   \frac{\delta G}{\delta v^{i}}
   \mathcal{J}^{ij}
   \frac{\delta H}{\delta v^{j}}
\nonumber
\\
&=
 G \left\{ F, H \right\}
+ F \left\{ G, H \right\},
\end{align}
where $H[v]$ is another arbitrary functional.
By using the antisymmetry, it is obvious to see
$ \{ F , G H \} = G \{ F, H \} + H \{ F, G \}$.

\subsubsection{Jacobi Identity}
\label{subsubsec:ProofLiePoissonBracketJacobiIdentity}

Let us recall that the elements of the Poisson tensor 
(\ref{eq:PoissonTensor}) 
have the form $\mathcal{J}^{ij} = [ \circ, J^{ij} ]$.
Moreover, $J^{ij}$ can be expressed as $J^{ij} = v^{k} a^{ij}_{k}$,
where
$a^{11}_{1} = a^{12}_{2} = a^{21}_{2} = a^{13}_{3}= a^{31}_{3}
= a^{24}_{3} = a^{42}_{3} = a^{14}_{4} = a^{41}_{4} = 1$
and 
$a^{11}_{2} = a^{14}_{3} = a^{41}_{3} = -2 M K / ( K_{0}^{2} r^{2} )^{2}
= f$,
and 
 $a^{ij}_{k} = 0$ otherwise.
%Note that this $f$ is the one defined in Eq.~(\ref{eq:f-definition}),
%not the one used in Eq.~(\ref{eq:functionalF}).

The Lie--Poisson bracket (\ref{eq:LiePoissonBracket1})
can be written as
\begin{align}
 \{ F, G\}
&= 
 \int \td V \,
 v^{k}
 a^{ij}_{k} 
 \left[
   \frac{\delta F}{\delta v^{i}} ,
   \frac{\delta G}{\delta v^{j}}
 \right]
\nonumber
\\
&=: 
 \left\langle
   v , 
   \left[\!\!\left[ 
      \frac{\delta F}{\delta v} ,
      \frac{\delta G}{\delta v}
   \right]\!\!\right]
 \right\rangle, 
\end{align}
where $\langle \, , \rangle$ denotes an inner product
and 
\begin{equation}
   \left[\!\!\left[ 
      \frac{\delta F}{\delta v} ,
      \frac{\delta G}{\delta v}
   \right]\!\!\right]_{k}
:=
 a^{ij}_{k} 
 \left[ 
   \frac{\delta F}{\delta v^{i}} ,
   \frac{\delta G}{\delta v^{j}}
 \right].
\end{equation}

A first variation of $\{ F , G \}$ can be calculated,
by using arbitrary functions $\tilde{v} = ( \tilde{v}^{k} )$ satisfying
the same boundary condition as $v$, as
\begin{align}
 \delta 
 \left\langle
   v , 
   \left[\!\!\left[ 
      \frac{\delta F}{\delta v} ,
      \frac{\delta G}{\delta v}
   \right]\!\!\right]
 \right\rangle
&=
 \int \td V \,
 \left(
 \tilde{v}^{k} 
   \left[\!\!\left[ 
      \frac{\delta F}{\delta v} ,
      \frac{\delta G}{\delta v}
   \right]\!\!\right]_{k}
 \right.
\nonumber
\\
& \hspace{2cm}
 \left.
 + v^{k} 
   a^{ij}_{k}
   \left(
     \left[
       \frac{\delta^{2} F}{\delta v^{i} \delta v^{\ell}} \tilde{v}^{\ell} ,
       \frac{\delta G}{\delta v^{j}}
     \right]
+
     \left[
       \frac{\delta F}{\delta v^{i}} ,
       \frac{\delta^{2} G}{\delta v^{j} \delta v^{\ell}} \tilde{v}^{\ell} 
     \right]
   \right)
  \right)
\nonumber
\\
&=
 \int \td V \,
 \left(
 \tilde{v}^{k} 
   \left[\!\!\left[ 
      \frac{\delta F}{\delta v} ,
      \frac{\delta G}{\delta v}
   \right]\!\!\right]_{k}
 \right.
\nonumber
\\
& \hspace*{2cm}
 \left.
 + \left(
     \frac{\delta^{2} F}{\delta v^{i} \delta v^{\ell}} \tilde{v}^{\ell}
     \left[
       \frac{\delta G}{\delta v^{j}} ,
       v^{k} a^{ij}_{k}
     \right]
+
     \frac{\delta^{2} G}{\delta v^{j} \delta v^{\ell}} \tilde{v}^{\ell} 
     \left[
       v^{k} a^{ij}_{k} ,
       \frac{\delta F}{\delta v^{i}}        
     \right]
   \right) 
 \right)
\nonumber
\\
&=
 \int \td V \,
 \tilde{v}^{\ell} 
 \left(
   \left[\!\!\left[ 
      \frac{\delta F}{\delta v} ,
      \frac{\delta G}{\delta v}
   \right]\!\!\right]_{\ell}
-  \frac{\delta^{2} F}{\delta v^{i} \delta v^{\ell}}
     \left[
       v^{k} a^{ij}_{k} ,
       \frac{\delta G}{\delta v^{j}} 
     \right]
+
     \frac{\delta^{2} G}{\delta v^{j} \delta v^{\ell}}
     \left[
       v^{k} a^{ij}_{k} ,
       \frac{\delta F}{\delta v^{i}}        
     \right]
   \right).
\end{align}
Therefore,
\begin{equation}
 \frac{\delta}{\delta v^{\ell}}
 \left\langle
   v , 
   \left[\!\!\left[ 
      \frac{\delta F}{\delta v} ,
      \frac{\delta G}{\delta v}
   \right]\!\!\right]
 \right\rangle
=
   \left[\!\!\left[ 
      \frac{\delta F}{\delta v} ,
      \frac{\delta G}{\delta v}
   \right]\!\!\right]_{\ell}
-   \frac{\delta^{2} F}{\delta v^{i} \delta v^{\ell}}
     \left[
       v^{k} a^{ij}_{k} , 
       \frac{\delta G}{\delta v^{j}}
     \right]
+
     \frac{\delta^{2} G}{\delta v^{j} \delta v^{\ell}}
     \left[
       v^{k} a^{ij}_{k} ,
       \frac{\delta F}{\delta v^{i}}        
     \right].
\end{equation}

Then, using also another arbitrary functional
$H[v]$,
\begin{align}
 \{ H , \{ F, G \}  \}
&=
 \left\langle
   v ,
   \left[\!\!\left[ 
      \frac{\delta H}{\delta v} ,
 \frac{\delta}{\delta v}
 \left\langle
   v , 
   \left[\!\!\left[ 
      \frac{\delta F}{\delta v} ,
      \frac{\delta G}{\delta v}
   \right]\!\!\right]
 \right\rangle
   \right]\!\!\right]
 \right\rangle
\nonumber
\\
&=
 \left\langle
   v,
   \left[\!\!\left[ 
   \frac{\delta H}{\delta v} ,
   \left[\!\!\left[ 
      \frac{\delta F}{\delta v} ,
      \frac{\delta G}{\delta v}
   \right]\!\!\right]
   \right]\!\!\right]
 \right\rangle
\nonumber
\\
& \hspace*{1cm}
+
 \int \td V \,
 v^{n}
 a^{m \ell}_{n} 
 \left[
   \frac{\delta H}{\delta v^{m}} ,
-   \frac{\delta^{2} F}{\delta v^{i} \delta v^{\ell}}
     \left[
       v^{k} a^{ij}_{k} , 
       \frac{\delta G}{\delta v^{j}}
     \right]
+
     \frac{\delta^{2} G}{\delta v^{j} \delta v^{\ell}}
     \left[
       v^{k} a^{ij}_{k} ,
       \frac{\delta F}{\delta v^{i}}        
     \right]   
 \right]
\nonumber
\\
&=
 \left\langle
   v,
   \left[\!\!\left[ 
   \frac{\delta H}{\delta v} ,
   \left[\!\!\left[ 
      \frac{\delta F}{\delta v} ,
      \frac{\delta G}{\delta v}
   \right]\!\!\right]
   \right]\!\!\right]
 \right\rangle
\nonumber
\\
& \hspace*{1cm}
+
 \int \td V \,
 \left(
 -  \frac{\delta^{2} F}{\delta v^{i} \delta v^{\ell}}
     \left[
       v^{k} a^{ij}_{k} ,
       \frac{\delta G}{\delta v^{j}}
     \right]
 \left[
 v^{n}
 a^{m \ell}_{n} ,
 \frac{\delta H}{\delta v^{m}} 
 \right]
 \right.
\nonumber
\\
& \hspace*{3cm}
\left.
+
 \frac{\delta^{2} G}{\delta v^{j} \delta v^{\ell}}
     \left[
       v^{k} a^{ij}_{k} ,
       \frac{\delta F}{\delta v^{i}}        
     \right]   
 \left[
  v^{n}
 a^{m \ell}_{n} ,
 \frac{\delta H}{\delta v^{m}}
 \right]
 \right).
\end{align}
Because 
$\delta^{2} F / ( \delta v^{i} \delta v^{\ell} )$
and
$\delta^{2} G / ( \delta v^{j} \delta v^{\ell} )$ 
are symmetric, 
we obtain
\begin{align}
&
  \{ H , \{ F, G \}  \}
+ \{ F , \{ G, H \}  \}
+ \{ G , \{ H, F \}  \}
\nonumber
\\
&
=
\left\langle
 v ,
   \left[\!\!\left[ 
   \frac{\delta H}{\delta v} ,
   \left[\!\!\left[ 
      \frac{\delta F}{\delta v} ,
      \frac{\delta G}{\delta v}
   \right]\!\!\right]
   \right]\!\!\right]
+   \left[\!\!\left[ 
   \frac{\delta F}{\delta v} ,
   \left[\!\!\left[ 
      \frac{\delta G}{\delta v} ,
      \frac{\delta H}{\delta v}
   \right]\!\!\right]
   \right]\!\!\right]
+   \left[\!\!\left[ 
   \frac{\delta G}{\delta v} ,
   \left[\!\!\left[ 
      \frac{\delta H}{\delta v} ,
      \frac{\delta F}{\delta v}
   \right]\!\!\right]
   \right]\!\!\right]
\right\rangle.
\label{eq:JacobiIdentityParingExpression}
\end{align}

If the second argument of
Eq.~(\ref{eq:JacobiIdentityParingExpression}) vanishes,
the Jacobi identity for the Lie--Poisson bracket 
(\ref{eq:LiePoissonBracket1}) is proven.
The explicit expression of the second argument is written as
\begin{align}
&
 \left(
    \left[\!\!\left[ 
   \frac{\delta H}{\delta v} ,
   \left[\!\!\left[ 
      \frac{\delta F}{\delta v} ,
      \frac{\delta G}{\delta v}
   \right]\!\!\right]
   \right]\!\!\right]
+   \left[\!\!\left[ 
   \frac{\delta F}{\delta v} ,
   \left[\!\!\left[ 
      \frac{\delta G}{\delta v} ,
      \frac{\delta H}{\delta v}
   \right]\!\!\right]
   \right]\!\!\right]
+   \left[\!\!\left[ 
   \frac{\delta G}{\delta v} ,
   \left[\!\!\left[ 
      \frac{\delta H}{\delta v} ,
      \frac{\delta F}{\delta v}
   \right]\!\!\right]
   \right]\!\!\right]
 \right)_{\ell}
\nonumber
\\
&=
  a^{km}_{\ell}
  \left[
  \frac{\delta H}{\delta v^{k}} , 
   \left[\!\!\left[ 
      \frac{\delta F}{\delta v} ,
      \frac{\delta G}{\delta v}
   \right]\!\!\right]_{m}
  \right]
+ a^{im}_{\ell}
  \left[
  \frac{\delta F}{\delta v^{i}} , 
   \left[\!\!\left[ 
      \frac{\delta G}{\delta v} ,
      \frac{\delta H}{\delta v}
   \right]\!\!\right]_{m}
  \right]
+ a^{jm}_{\ell}
  \left[
  \frac{\delta G}{\delta v^{j}} ,
   \left[\!\!\left[ 
      \frac{\delta H}{\delta v} ,
      \frac{\delta F}{\delta v}
   \right]\!\!\right]_{m}
  \right]
\nonumber
\\
&=
  a^{km}_{\ell}
  \left[
  \frac{\delta H}{\delta v^{k}} , 
   a^{ij}_{m}
   \left[
      \frac{\delta F}{\delta v^{i}} ,
      \frac{\delta G}{\delta v^{j}}
   \right]
  \right]
+ a^{im}_{\ell}
  \left[
  \frac{\delta F}{\delta v^{i}} , 
  a^{jk}_{m}
  \left[
      \frac{\delta G}{\delta v^{j}} ,
      \frac{\delta H}{\delta v^{k}}
  \right]
  \right]
+ a^{jm}_{\ell}
  \left[
  \frac{\delta G}{\delta v^{j}} ,
  a^{ki}_{m}
  \left[
      \frac{\delta H}{\delta v^{k}} ,
      \frac{\delta F}{\delta v^{i}}
  \right]
  \right]
\nonumber
\\
&=
  a^{km}_{\ell} a^{ij}_{m}
  \left[
  \frac{\delta H}{\delta v^{k}} , 
   \left[
      \frac{\delta F}{\delta v^{i}} ,
      \frac{\delta G}{\delta v^{j}}
   \right]
  \right]
+ a^{im}_{\ell} a^{jk}_{m}
  \left[
  \frac{\delta F}{\delta v^{i}} , 
  \left[
      \frac{\delta G}{\delta v^{j}} ,
      \frac{\delta H}{\delta v^{k}}
  \right]
  \right]
+ a^{jm}_{\ell} a^{ki}_{m}
  \left[
  \frac{\delta G}{\delta v^{j}} ,
  \left[
      \frac{\delta H}{\delta v^{k}} ,
      \frac{\delta F}{\delta v^{i}}
  \right]
  \right]
\nonumber
\\
& \hspace*{0.5cm}
+ a^{km}_{\ell}
  \left[
  \frac{\delta H}{\delta v^{k}} , 
   a^{ij}_{m}
  \right]
   \left[
      \frac{\delta F}{\delta v^{i}} ,
      \frac{\delta G}{\delta v^{j}}
   \right]
+ a^{im}_{\ell}
  \left[
  \frac{\delta F}{\delta v^{i}} , 
  a^{jk}_{m}
  \right]
  \left[
      \frac{\delta G}{\delta v^{j}} ,
      \frac{\delta H}{\delta v^{k}}
  \right]
+ a^{jm}_{\ell}
  \left[
  \frac{\delta G}{\delta v^{j}} ,
  a^{ki}_{m}
  \right]
  \left[
      \frac{\delta H}{\delta v^{k}} ,
      \frac{\delta F}{\delta v^{i}}
  \right]
\nonumber
\\
&=
  a^{km}_{\ell} a^{ij}_{m}
  \left[
  \frac{\delta H}{\delta v^{k}} , 
   \left[
      \frac{\delta F}{\delta v^{i}} ,
      \frac{\delta G}{\delta v^{j}}
   \right]
  \right]
+ a^{im}_{\ell} a^{jk}_{m}
  \left[
  \frac{\delta F}{\delta v^{i}} , 
  \left[
      \frac{\delta G}{\delta v^{j}} ,
      \frac{\delta H}{\delta v^{k}}
  \right]
  \right]
+ a^{jm}_{\ell} a^{ki}_{m}
  \left[
  \frac{\delta G}{\delta v^{j}} ,
  \left[
      \frac{\delta H}{\delta v^{k}} ,
      \frac{\delta F}{\delta v^{i}}
  \right]
  \right]
\nonumber
\\
& \hspace*{0.5cm}
+ \frac{1}{(K r)^{2}}
  \left(
    \left(
       a^{jm}_{\ell} \frac{\td a^{ki}_{m}}{\td r}
     - a^{km}_{\ell} \frac{\td a^{ij}_{m}}{\td r}
    \right)
    \frac{\pd}{\pd r} \left( \frac{\delta F}{\delta v^{i}} \right)
    \frac{\pd}{\pd \alpha} \left( \frac{\delta G}{\delta v^{j}} \right)
    \frac{\pd}{\pd \alpha} \left( \frac{\delta H}{\delta v^{k}} \right)
  \right.
\nonumber
\\
& \hspace*{2cm}
  \left.
+   \left(
       a^{km}_{\ell} \frac{\td a^{ij}_{m}}{\td r}
     - a^{im}_{\ell} \frac{\td a^{jk}_{m}}{\td r}
    \right)
    \frac{\pd}{\pd r} \left( \frac{\delta G}{\delta v^{j}} \right)    
    \frac{\pd}{\pd \alpha} \left( \frac{\delta H}{\delta v^{k}} \right)
    \frac{\pd}{\pd \alpha} \left( \frac{\delta F}{\delta v^{i}} \right)
  \right.
\nonumber
\\
& \hspace*{2cm}
  \left.
+   \left(
       a^{im}_{\ell} \frac{\td a^{jk}_{m}}{\td r}
     - a^{jm}_{\ell} \frac{\td a^{ki}_{m}}{\td r}
    \right)
    \frac{\pd}{\pd r} \left( \frac{\delta H}{\delta v^{k}} \right)
    \frac{\pd}{\pd \alpha} \left( \frac{\delta F}{\delta v^{i}} \right)
    \frac{\pd}{\pd \alpha} \left( \frac{\delta G}{\delta v^{j}} \right)
  \right).
\label{eq:secondArgumentJacobiIdeneityParingExpression}
\end{align}
Therefore, if the following two conditions are satisfied 
for all combinations of $i$, $j$, $k$ and $\ell$, 
Eq.~(\ref{eq:secondArgumentJacobiIdeneityParingExpression}) vanishes:
\begin{alignat}{2}
& \text{Condition (I):}
&\hspace{1cm}&
  a^{km}_{\ell} a^{ij}_{m}
= a^{im}_{\ell} a^{jk}_{m} 
= a^{jm}_{\ell} a^{ki}_{m},
\label{eq:JacobiIdentityConditionI}
\\
& \text{Condition (II):}
&\hspace{1cm}& 
  a^{km}_{\ell} \frac{\td a^{ij}_{m}}{\td r}
= a^{im}_{\ell} \frac{\td a^{jk}_{m}}{\td r}
= a^{jm}_{\ell} \frac{\td a^{ki}_{m}}{\td r}.
\label{eq:JacobiIdentityConditionII}
\end{alignat}

Table~\ref{table:JacobiIdentityConditionI} summarizes 
the possible combinations of $(\ell, i, j, k)$ that give
nonzero contribution to Condition (I).
We observe that Condition (I) is satisfied for all the combinations.

\begin{table}[h]
 \centering
 \caption{Condition (I).}
 \begin{tabular*}{0.7\textwidth}[t]{@{\extracolsep{\fill}} ccccccc}
  $\ell$ & $i$ & $j$ & $k$ & 
  $a^{km}_{2} a^{ij}_{m}$ &
  $a^{im}_{2} a^{jk}_{m}$ & 
  $a^{jm}_{2} a^{ki}_{m}$  \\
  \hline \hline 
  $1$ & $1$ & $1$ & $1$ & $1$  & $1$  & $1$  \\
  \hline
  $2$ & $1$ & $1$ & $1$ & $2f$  & $2f$  & $2f$  \\
  $2$ & $1$ & $1$ & $2$ & $1$  & $1$  & $1$  \\
  $2$ & $1$ & $2$ & $1$ & $1$  & $1$  & $1$  \\
  $2$ & $2$ & $1$ & $1$ & $1$  & $1$  & $1$  \\
  \hline
  $3$ & $1$ & $1$ & $3$ & $1$  & $1$  & $1$  \\
  $3$ & $1$ & $3$ & $1$ & $1$  & $1$  & $1$  \\
  $3$ & $3$ & $1$ & $1$ & $1$  & $1$  & $1$  \\
  $3$ & $1$ & $2$ & $4$ & $1$  & $1$  & $1$  \\
  $3$ & $1$ & $4$ & $2$ & $1$  & $1$  & $1$  \\
  $3$ & $2$ & $1$ & $4$ & $1$  & $1$  & $1$  \\
  $3$ & $2$ & $4$ & $1$ & $1$  & $1$  & $1$  \\
  $3$ & $4$ & $1$ & $2$ & $1$  & $1$  & $1$  \\
  $3$ & $4$ & $2$ & $1$ & $1$  & $1$  & $1$  \\
  $3$ & $1$ & $1$ & $4$ & $2f$  & $2f$  & $2f$  \\
  $3$ & $1$ & $4$ & $1$ & $1$  & $1$  & $1$  \\
  $3$ & $4$ & $1$ & $2$ & $1$  & $1$  & $1$  \\
  \hline
  $4$ & $1$ & $1$ & $4$ & $1$  & $1$  & $1$  \\
  $4$ & $1$ & $4$ & $1$ & $1$  & $1$  & $1$  \\
  $4$ & $4$ & $1$ & $1$ & $1$  & $1$  & $1$  
 \end{tabular*}
 \label{table:JacobiIdentityConditionI}
\end{table}

Similarly, Table~\ref{table:JacobiIdentityConditionII} summarizes 
the possible combinations of $(\ell, i, j, k)$ that give
nonzero contribution to Condition (II).
For Condition (II), only 
$\td a^{11}_{2} / \td r$,
$\td a^{14}_{3} / \td r$, and
$\td a^{41}_{3} / \td r$ remain nonzero.
Therefore, nonzero contributions arise for $\ell = 2$ and $3$,
and no contribution appears for $\ell = 1$ and $4$.
The prime denotes a derivative with respect to $r$.
In all possible cases shown in 
Table~\ref{table:JacobiIdentityConditionII},
Condition (II) holds.

\begin{table}[h]
 \centering
 \caption{Condition (II).}
 \begin{tabular*}{0.7\textwidth}[t]{@{\extracolsep{\fill}} ccccccc}
  $\ell$ & $i$ & $j$ & $k$ & 
  $a^{km}_{2} \td a^{ij}_{m} / \td r$ &
  $a^{im}_{2} \td a^{jk}_{m} / \td r$ & 
  $a^{jm}_{2} \td a^{ki}_{m} / \td r$ \\
  \hline \hline 
  $2$ & $1$ & $1$ & $1$ & $f^{\pr}$  & $f^{\pr}$  & $f^{\pr}$ \\
  \hline
  $3$ & $1$ & $1$ & $4$ & $f^{\pr}$  & $f^{\pr}$  & $f^{\pr}$  \\
  $3$ & $1$ & $4$ & $1$ & $f^{\pr}$  & $f^{\pr}$  & $f^{\pr}$  \\
  $3$ & $4$ & $1$ & $1$ & $f^{\pr}$  & $f^{\pr}$  & $f^{\pr}$ 
 \end{tabular*}
 \label{table:JacobiIdentityConditionII}
\end{table}

With these examinations, we conclude that 
the Lie--Poisson bracket (\ref{eq:LiePoissonBracket1}) 
satisfies 
the Jacobi identity (\ref{eq:LiePoissonBracketJacobiIdentity}).

\subsection{Casimir invariants}

At the last of Sec~\ref{sec:HamiltonianStructure}, 
the Casimir invariants $C[v]$ are found by integrating 
$\mathcal{J}^{ij} \delta C / \delta v^{j} = 0$ for $i = 1, 2, 3, 4$ 
as
\begin{align}
 C[v]
&=
 \int \td V \,  \left(
    U F_{1}(\psi) 
    + \left(
        u_{h} F_{1}^{\pr}(\psi) + F_{2}^{\pr}(\psi) 
      \right)
      \left(
        B_{h}^{\star} + f(r) \psi
      \right)
    - f(r) F_{2}(\psi) 
    + u_{h} F_{3}(\psi)
    + F_{4}(\psi)
                \right)
\label{eq:Casimir1}
\\
&=
 \int \td V \,  \left(
    \vu \cdot \vB F_{1}^{\pr}(\psi) 
    + | \vh |^{2} B_{h} F_{2}^{\pr}(\psi) 
    - f(r) F_{2}(\psi) 
    + u_{h} F_{3}(\psi)
    + F_{4}(\psi)
                \right),
\label{eq:Casimir2}
\end{align}
where $F_{i}(\psi) \, (i = 1, 2, 3, 4)$ are arbitrary functions of
$\psi$.
Note that the Casimir invariants Eq.~(\ref{eq:Casimir1}) or
(\ref{eq:Casimir2}) cannot be derived from the Casimir invariants of the
full set of ideal MHD since the ideal MHD has only two Casimir
invariants; the magnetic helicity and the cross helicity.

Although the present model is applicable to only single-helicity
dynamics of plasmas in cylindrical geometry, 
it can be used for 
equilibrium and stability analysis of helically symmetric systems 
through the energy--Casimir method
based on the Hamiltonian structure presented in this Section.
This will be reported elsewhere.
Such a study might be of interest in relations to quasi-single-helicity
states observed in reversed field pinches.

An extension to a toroidal geometry may be interesting, however, 
it seems difficult to find an incompressible vector field 
such as $\vh$ in the present paper.  We may need some approximations such as
large aspect ratio.

\section{Applications}
\label{sec:application}

In this Section~\ref{sec:application},
we apply the derived model to two specific problems both related to the
MRI.
The first application is the local dispersion relation of axisymmetric
MRI, which will be described in Sec.~\ref{subsec:localDispersionRelation}.
The second application is the linear stability analyses 
and the calculation of mode energy in a setting of 
an accretion disk, which will be presented in 
Sec.~\ref{subsec:negativeEnergymode}.  
Both applications show that the derived model successfully reproduce the
results of previous studies.

\subsection{Local dispersion relation of axisymmetric MRI}
\label{subsec:localDispersionRelation}

Suppose that a cylindrically symmetric equilibrium with a 
velocity field $\vu_{0}$ and a magnetic field $\vB_{0}$ is given by
\begin{align}
 \vu_{0} 
&=
 u_{\theta 0}(r) \hvtheta + u_{z 0}(r) \hrvz,
 \label{eq:cylindricallySymmetricEquilibriumFlowVelocityGeneral}
\\
 \vB_{0} 
&=
 B_{\theta 0}(r) \hvtheta + B_{z 0}(r) \hrvz.
 \label{eq:cylindricallySymmetricEquilibriumMagneticFieldGeneral}
\end{align}
By comparing these expressions with 
Eqs.~(\ref{eq:vu}) and (\ref{eq:vB}), 
we obtain the following relations
\begin{align}
 \frac{\td \vphi_{0}}{\td r}
&=
 M u_{\theta 0} + K r u_{z 0},
 \label{eq:phi0pGeneral}
\\
 u_{h 0}
&=
 -K r u_{\theta 0} + M u_{z 0},
 \label{eq:uh0General}
\\
 \frac{\td \psi_{0}}{\td r}
&=
 - M B_{\theta 0} - K r B_{z 0},
 \label{eq:psi0pGeneral}
\\
 B_{h 0}
&=
 -K r B_{\theta 0} + M B_{z 0}.
 \label{eq:Bh0General}
\end{align}

Let us consider an equilibrium with 
$u_{\theta 0} = r \varOmega(r)$, $u_{z 0} = 0$,
$B_{\theta 0} = 0$, and $B_{z 0} = B_{0} = \mathrm{const}$.
Then we have
\begin{align}
 \vphi_{0}^{\pr}
&=
 0,
\\
 u_{h 0} 
&=
 - K r^{2} \varOmega,
\\
 \psi_{0}^{\pr}
&=
 -K r B_{0},
\\
 B_{h 0}
&=
 0.
\end{align}
Here the prime denotes a derivative with respect to $r$.

For deriving a local dispersion relation of axisymmetric MRI, 
let us set $M=0$, 
and drop $\pd / \pd r$ operating on perturbed quantities.
This is justified under the WKB approximation 
explained  in IV.A of Ref.~\onlinecite{B-H-1998} for example, 
where the perturbation is assumed to behave as 
$\rme^{\rmi ( \vk \cdot \vr - \omega t)}$
with $\vk = k \hrvz$ and $k r \gg 1$.  Then, the terms with
 $\pd / \pd r$ become higher order. 
From the linearized Eqs.~(\ref{eq:h.vorticityEq}), 
(\ref{eq:h.eqMotion}), (\ref{eq:h.OhmLaw}), and (\ref{eq:h.inductionEq})
about this equilibrium, 
we obtain 
\begin{equation}
\begin{pmatrix}
-\frac{\rmi \ell}{r} K ( r^{2} \varOmega )^{\pr}
& 
\rmi \omega  
& 
0
&  
\rmi \ell K B_{0}
\\
\rmi \omega 
 \left(
   - \frac{\ell^{2}}{r^{2}}
 \right)
&
\frac{2 \rmi \ell \varOmega}{K r^{2}}
& 
\frac{\rmi \ell^{3} K B_{0}}{r^{2}}
&
0
\\
0
&
\rmi \ell K B_{0}
& 
- \rmi \ell K r \varOmega^{\pr}
&
\rmi \omega
\\
- \rmi \ell K B_{0}
&
0
&
\rmi \omega  
&
0
\end{pmatrix}
\begin{pmatrix}
 \tilde{\vphi}
\\
\tilde{u}_{h}
\\
\tilde{\psi}
\\
\tilde{B}_{h}
\end{pmatrix}
=
\begin{pmatrix}
0
\\
0
\\
0
\\
0
\end{pmatrix},
\end{equation}
where $\tilde{\vphi}$, 
$\tilde{u}_{h}$, $\tilde{\psi}$, and $\tilde{B}_{h}$ 
are the perturbed quantities.

Setting the determinant of the $4 \times 4$ matrix on the left-hand side
zero,
we obtain 
\begin{equation}
 \omega^{2}
=
  \omega_{\mathrm{A}}^{2}  
 + \frac{1}{2} \frac{\td \varOmega^{2}}{ \td \ln r}
 + 2 \varOmega^{2}
 \pm \sqrt{
   \left(
    \frac{1}{2} \frac{\td \varOmega^{2}}{ \td \ln r}
     + 2 \varOmega^{2}
   \right)^{2}
  + 4 \omega_{\mathrm{A}}^{2} \varOmega^{2}
     },
\label{eq:localDispersionMRIomega2}
\end{equation}
where the Alfv\'en frequency $\omega_{\mathrm{A}}$ is given by
\begin{equation}
 \omega_{\mathrm{A}}
:=
 \sqrt{ \ell^{2} K^{2} B_{0}^{2} }.
\label{eq:AlfvenFrequency}
\end{equation}
Equation~(\ref{eq:localDispersionMRIomega2}) agrees with, 
for example, the 
dispersion relation Eq.~(111) of Ref.~\onlinecite{B-H-1998}.
If we assume 
$\dis{
\left|
  \frac{1}{2} \frac{\td \varOmega^{2}}{ \td \ln r}
 + 2 \varOmega^{2}
\right|
\gg 
\left|
 2 \omega_{\mathrm{A}} \varOmega
\right|
}$,
which means that the magnetic field is sufficiently weak so that the
Alfv\'en frequency $\omega_{\mathrm{A}}$ is small enough compared to the
plasma rotation frequency,
the square root term of Eq.~(\ref{eq:localDispersionMRIomega2})
can be approximated to yield
\begin{equation}
 \omega^{2}
\simeq
   \omega_{\mathrm{A}}^{2}  
 + \frac{1}{2} \frac{\td \varOmega^{2}}{ \td \ln r}
 + 2 \varOmega^{2}
 \pm 
 \left|
  \frac{1}{2} \frac{\td \varOmega^{2}}{ \td \ln r}
 + 2 \varOmega^{2}
 \right|
\pm 
  \frac{ 2 \omega_{\mathrm{A}}^{2} \varOmega^{2} }
       { \left| 
         \frac{1}{2} \frac{\td \varOmega^{2}}{ \td \ln r}
         + 2 \varOmega^{2} 
         \right|
       }.
\end{equation}
Then, we find that $\omega^{2} < 0$ when 
\begin{equation}
 - 4 \varOmega^{2}
 <
 \frac{\td \varOmega^{2}}{ \td \ln r}
 < 
 0.
\label{eq:localDispersionMRI1}
\end{equation}
By introducing the epicyclic frequency $\kappa$ through
\begin{equation}
 \kappa^{2}
:=
 \frac{1}{r^{3}}
 \frac{\td ( r^{4} \varOmega^{4} )}{ \td r},
\label{eq:epicyclicFrequency}
\end{equation}
the instability criterion (\ref{eq:localDispersionMRI1}) can be written
as 
\begin{equation}
 0 < \kappa^{2} < 4 \varOmega^{2}.
\label{eq:localDispersionMRI2}
\end{equation}
This agrees with the known condition for the instability.

\subsection{Negative Energy modes in an accretion disk}
\label{subsec:negativeEnergymode}

It was shown that negative energy modes can be found in an accretion
disk system in Ref.~\onlinecite{Khalzov-2008}.  They obtained the eigenmodes by
using the Frieman--Rotenberg equation\cite{F-R-1960}.
Here we obtain similar results by using the model derived 
in Sec.~\ref{sec:derivationGoverningEquations}.

The settings of the problem were taken to be the same as
Ref.~\onlinecite{Khalzov-2008}. 
A plasma fills an annular region between coaxial
cylindrical surfaces with their radii $r_{1}$ and $r_{2}$.  
The outer
radius is assumed to be $r_{2} = 5 r_{1}$.  
The length of the cylinder was taken to be $2 \pi R_{0} = 4 r_{2} / 5$.
The plasma rotation velocity at the equilibrium is 
$\vu_{0} = r \Omega(r) \hvtheta$ with 
$\varOmega(r) = \varOmega_{1} r_{1}^{2} / r^{2}$.
A uniform magnetic field $\vB_{0} = B_{0} \hrvz$ is assumed.  

The governing equations 
(\ref{eq:h.vorticityEq}), 
(\ref{eq:h.eqMotion}), (\ref{eq:h.OhmLaw}), and (\ref{eq:h.inductionEq})
were linearized about the equilibrium.
Assuming the perturbed quantities behave as
$\rme^{\rmi ( \ell K \alpha - \omega t )}$, 
the linearized equations become
\begin{align}
 - \rmi \omega
   \mathcal{L} \tilde{\vphi}
&= 
 2 \rmi \frac{\ell}{r} K 
 \frac{- K r u_{h0} + M \vphi_{0}^{\pr} }{( K_{0}^{2} r^{2} )^{2}} 
 \tilde{u}_{h}
-\frac{\rmi \ell}{r}
 \left(
   \vphi_{0}^{\pr}
   \mathcal{L}
  -\left(
   \mathcal{L}_{0} \vphi_{0}
   \right)^{\pr}
  +2 M K
   \left(
    \frac{u_{h0}}{( K_{0}^{2} r^{2} )^{2}} 
   \right)^{\pr}
 \right)
 \tilde{\vphi}
\nonumber
\\
& \quad
+ 2 \rmi \frac{\ell}{r} K 
  \frac{ K r B_{h0} + M \psi_{0}^{\pr} }{( K_{0}^{2} r^{2} )^{2}}
  \tilde{B}_{h}
+\frac{\rmi \ell}{r}
 \left(
   \psi_{0}^{\pr}
   \mathcal{L} 
  -\left(
   \mathcal{L}_{0} \psi_{0}
   \right)^{\pr}
  -2 M K
   \left(
    \frac{B_{h0}}{( K_{0}^{2} r^{2} )^{2}}  
   \right)^{\pr}
 \right)
 \tilde{\psi},
\label{eq:h.vorticityEqLinear}
\\
  -\rmi \omega \tilde{u}_{h}
&=
 \frac{\rmi \ell}{r}
 \left(
  u_{h0}^{\pr} \tilde{\vphi} - \vphi_{0}^{\pr} \tilde{u}_{h}
+ B_{h0}^{\pr} \tilde{\psi} - \psi_{0}^{\pr} \tilde{B}_{h}
 \right),
\label{eq:h.eqMotionLinear}
\\
 -\rmi \omega \tilde{\psi}
&=
 \frac{\rmi \ell}{r}
 \left(
  \psi_{0}^{\pr} \tilde{\vphi}
 -\vphi_{0}^{\pr} \tilde{\psi}
 \right)
\label{eq:h.OhmLawLinear}
\\
 -\rmi \omega \tilde{B}_{h}
&=
 - \frac{\rmi \ell}{r}
   \psi_{0}^{\pr} \tilde{u}_{h}
 + \frac{\rmi \ell}{r}
   \left(
      K_{0}^{2} r^{2}
      \left(
        \frac{u_{h0}}{K_{0}^{2} r^{2}} 
      \right)^{\pr}
    + \frac{2 M K}{K_{0}^{2} r^{2}} 
      \vphi_{0}^{\pr}
   \right)
   \tilde{\psi}
\nonumber
\\
& \quad
 - \frac{\rmi \ell}{r}
   \vphi_{0}^{\pr} \tilde{B}_{h}
 + \frac{\rmi \ell}{r}
   \left(
      K_{0}^{2} r^{2}
      \left(
        \frac{B_{h0}}{K_{0}^{2} r^{2}} 
      \right)^{\pr}
    - \frac{2 M K}{K_{0}^{2} r^{2}} 
      \psi_{0}^{\pr}
   \right)
   \tilde{\vphi}
\label{eq:h.inductionEqLinear}
\end{align}
where 
each equilibrium quantity is written with a subscript $_{0}$,
the prime denotes the $r$ derivative, and 
\begin{equation}
 \mathcal{L}_{0}
:=
    \frac{1}{K r} 
   \frac{\td}{\td r} 
    \left( \frac{K r}{K_{0}^{2} r^{2}} \frac{\td }{\td r} \right)
\end{equation}
operates on equilibrium quantities.
Equations~(\ref{eq:h.eqMotionLinear}), (\ref{eq:h.OhmLawLinear})
 and (\ref{eq:h.inductionEqLinear})
are algebraic equations for the perturbed quantities. 
Therefore, $\tilde{u}_{h}$, $\tilde{\psi}$, and $\tilde{B}_{h}$
can be expressed by $\tilde{\vphi}$ by using these three equations.
Substituting 
$\tilde{u}_{h}$, $\tilde{\psi}$, and $\tilde{B}_{h}$
into Eq.~(\ref{eq:h.vorticityEqLinear}),
we obtain a second-order ordinary differential equation for 
$\tilde{\vphi}$ since $\mathcal{L}$ is 
a second-order differential operator in $r$.
Then we adopt boundary conditions
$\tilde{\vphi}(r_{1}) = \tilde{\vphi}(r_{2}) = 0$.
%Note that the linearized equations can be summarized in a second-order
%ordinary differential equation for $\tilde{\vphi}$, so that 
%the boundary conditions mentioned above are necessary and sufficient.
%The time dependence of the perturbed quantities were assumed to be 
%$\rme^{-\rmi \omega t}$.  
Here, however, we solved 
Eqs.~(\ref{eq:h.vorticityEqLinear})--(\ref{eq:h.inductionEqLinear})
by a finite element method with first-order elements, 
instead of solving the second-order ordinary differential equation
summarized in $\tilde{\vphi}$. 
Eigenvalues of the resultant matrix,
generated by discretization by the finite element method, 
 were calculated by Lapack.

\begin{figure}[h]
  \begin{minipage}[t]{0.45\textwidth}
    \centering
    \includegraphics[width=\textwidth]{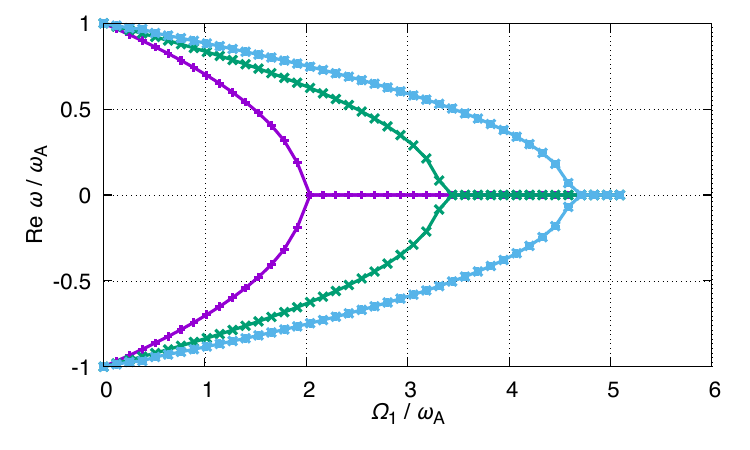}
    \subcaption{$\Re \, \omega / \omega_{\mathrm{A}}$.}
    \label{subfig:Omg1-omg_r-M0}
  \end{minipage}
  \begin{minipage}[t]{0.45\textwidth}
    \centering
    \includegraphics[width=\textwidth]{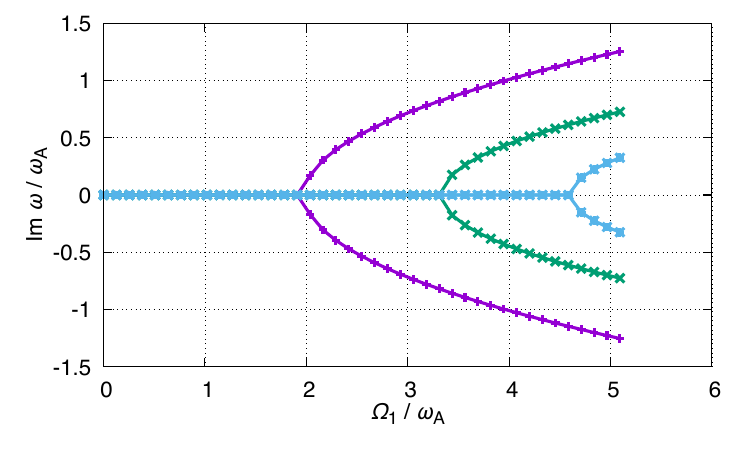}
    \subcaption{$\Im \, \omega / \omega_{\mathrm{A}}$.}
    \label{subfig:Omg1-omg_i-M0}
  \end{minipage}
  \begin{minipage}[t]{0.45\textwidth}
    \centering
    \includegraphics[width=\textwidth]{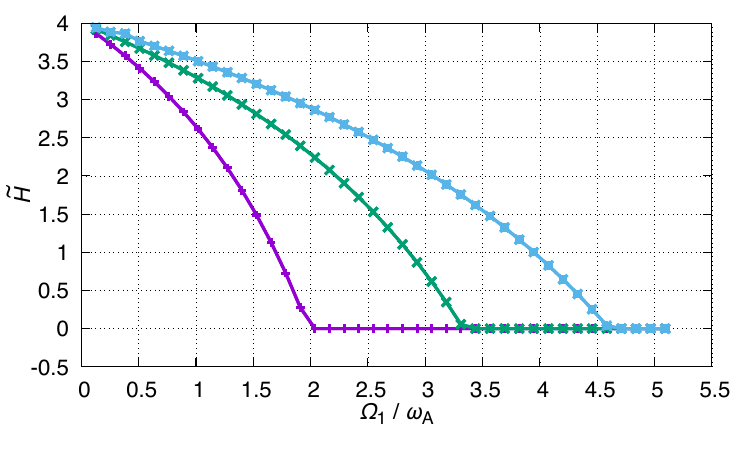}
    \subcaption{$\tilde{H}$.}
    \label{subfig:Omg1-ene-M0}
  \end{minipage}
  \caption{Eigenvalues and perturbed energy of axisymmetric modes with
 $M = 0$, $N = 1$, and $\ell = 1$.}
  \label{fig:M0}
\end{figure}

Figures~\ref{fig:M0}\subref{subfig:Omg1-omg_r-M0}
and \ref{fig:M0}\subref{subfig:Omg1-omg_i-M0}
show real and imaginary parts of the  eigenvalues 
as functions of $\varOmega_{1}$ for $M = 0$, $N = 1$
and $\ell = 1$.  The frequencies are normalized by the Alfv\'en
frequency $\omega_{\mathrm{A}}$.  
Three pairs of eigenvalues with the first to the third smallest
magnitudes of their real parts are plotted for each $\varOmega_{1}$.  
As $\varOmega_{1}$ is increased, the magnitudes of the real frequencies
become smaller, and the modes become unstable above the threshold values
of $\varOmega_{1}$.
The pair of modes that have the smallest real parts of $\omega$ 
become unstable first as $\varOmega_{1}$ is increased,
and have largest growth and damping rates, 
as plotted by purple markers and curves in 
Figs.~\ref{fig:M0}\subref{subfig:Omg1-omg_r-M0}
and \ref{fig:M0}\subref{subfig:Omg1-omg_i-M0}.

Figure~\ref{fig:M0}\subref{subfig:Omg1-ene-M0} shows the normalized
energy $\tilde{H}$ of the eigenmodes.  
Energy of an eigenmode was calculated on the basis of
Ref.~\onlinecite{Hirota-2008}
and normalized as Eq.~(\ref{eq:normalizedModeEnergy}).
The details of the perturbed energy calculation are described in
Appendix~\ref{sec:energyCalculation}. 
Note that this Appendix~\ref{sec:energyCalculation} explains only the
energy calculation of an eigenmode.  Energy calculation including
Alfv\'en continua was formulated in Ref.~\onlinecite{Hirota-2008-PoP}.

As we observe from Fig.~\ref{fig:M0}\subref{subfig:Omg1-ene-M0}, 
the mode energy decreases as $\varOmega_{1}$ is increased, and becomes
zero at the threshold values 
of $\varOmega_{1}$ for instability.

\begin{figure}[h]
  \begin{minipage}[t]{0.45\textwidth}
    \centering
    \includegraphics[width=\textwidth]{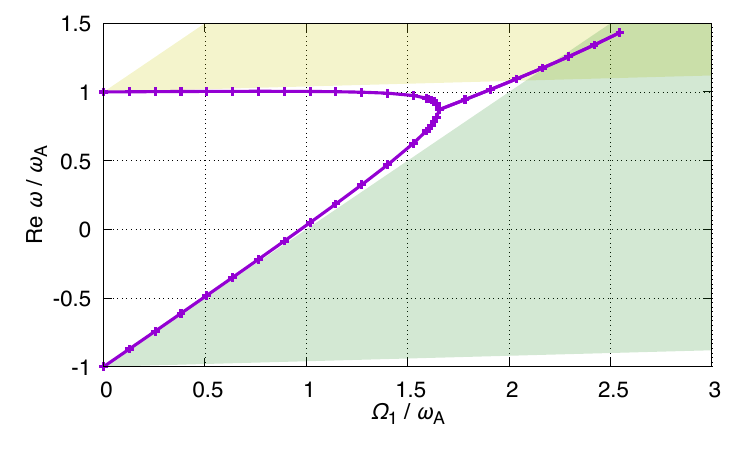}
    \subcaption{$\Re \, \omega / \omega_{\mathrm{A}}$.}
    \label{subfig:Omg1-omg_r-M1}
  \end{minipage}
  \begin{minipage}[t]{0.45\textwidth}
    \centering
    \includegraphics[width=\textwidth]{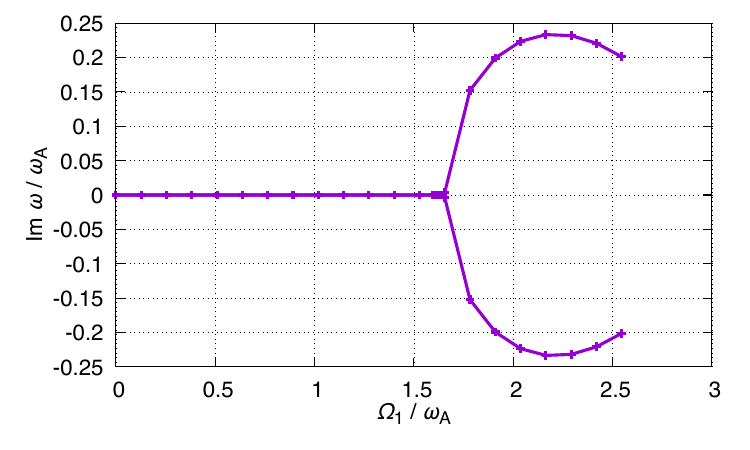}
    \subcaption{$\Im \, \omega / \omega_{\mathrm{A}}$.}
    \label{subfig:Omg1-omg_i-M1}
  \end{minipage}
  \begin{minipage}[t]{0.45\textwidth}
    \centering
    \includegraphics[width=\textwidth]{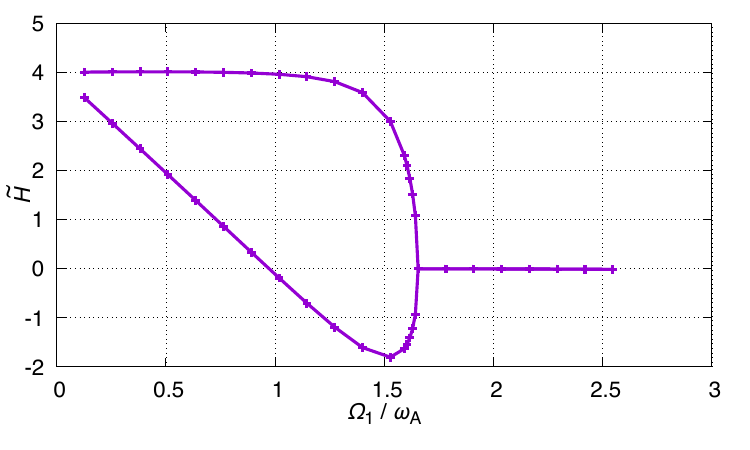}
    \subcaption{$\tilde{H}$.}
    \label{subfig:Omg1-ene-M1}
  \end{minipage}
  \caption{Eigenvalues and perturbed energy of non-axisymmetric modes
 with $M = 1$, $N = 1$, and $\ell = 1$.}
  \label{fig:M1}
\end{figure}

Similarly, the eigenvalues and the mode energy are plotted  
in Fig.~\ref{fig:M1} for $M = 1$, $N = 1$ and $\ell = 1$.
Only a pair of eigenmodes is shown, which is clearly in the frequency
range between the two Alfv\'en continua shaded by light yellow and 
green colors.  
The frequency ranges of the Alfv\'en continua are given by
\begin{equation}
 \frac{\omega}{\omega_{\mathrm{A}}}
 = \frac{\ell}{r} ( \pm \psi_{0}^{\pr} + \vphi_{0}^{\pr} )
 \label{eq:continuum-normalized}
\end{equation}
with
$\vphi_{0}^{\pr} = M r \varOmega(r) = M r_{1}^{2} \varOmega_{1} / r$
and
$\psi_{0}^{\pr} = - K r B_{0}$.

Note that we could obtain more branches of eigenmodes having 
positive real frequencies at $\varOmega_{1}$ smaller than the
instability threshold,
however, it was difficult to obtain branches of 
eigenmodes having 
negative real frequencies.
They seem to be in the Alfv\'en continuum
with the negative sign in Eq.~(\ref{eq:continuum-normalized}),
and the numerical results that might correspond to those branches were
unreliable numerically.

Reference~\onlinecite{Khalzov-2008} showed more pairs of eigenvalues.
According to their result, 
the magnitudes of $\Re\, \omega / \omega_{\mathrm{A}}$ of the second pair
are larger than those of the first pair, and the third pair has 
larger magnitudes of $\Re\, \omega / \omega_{\mathrm{A}}$ than those of
the second pair before $\Re\, \omega / \omega_{\mathrm{A}}$
changes the sign as  $\varOmega_{1}$ is increased.
Since the instability occurs when positive and negative energy modes
collide, 
the pair of eigenvalues shown in Fig.~\ref{fig:M1},
which collide first as $\varOmega_{1}$ is increased, are the most 
unstable modes.

In Fig.~\ref{fig:M1}, 
the eigenmode with negative real frequency at $\varOmega_{1} = 0$
changes its real frequency to be positive at 
$\varOmega_{1} / \omega_{\mathrm{A}} \simeq 1$,
and becomes a negative energy mode.  
The imaginary part is still zero in the range
$1 \lesssim \varOmega_{1} / \omega_{\mathrm{A}} \lesssim 1.6$,
the mode is a stable oscillation.  

The two eigenvalues with positive and negative energies 
collide at  
$\Omega_{1} / \omega_{\mathrm{A}} \simeq 1.6$, and become a pair of
eigenvalues of exponentially growing and the corresponding damping
modes.
This is a Hamiltonian Hopf (Krein) bifurcation.  
The energy of the unstable eigenmodes is zero.

\section{Conclusions}
\label{sec:conclusion}

We derived a four-field reduced model of single-helicity, incompressible MHD
in cylindrical geometry.  
We clarified the Hamiltonian structure of the model,
where the appropriate noncanonical variables were found,
and the corresponding Lie--Poisson bracket were constructed.
The Lie--Poisson bracket was proven to satisfy 
the antisymmetry, the Leibniz rule, and the Jacobi identity.
We found the Casimir invariants including four arbitrary functions.

In the present model, the direction of symmetry is defined by the vector
field $\vh$,
which has nonzero helicity.  This is different from systems with
axisymmetry or translational symmetry in the axial direction of the
cylindrical plasma.  The Hamiltonian structure of the present model was
clarified even though the helicity of $\vh$ remains finite. 

The model was applied to derive the well-known local dispersion relation
of the axisymmetric magnetorotational instability, and was shown to
successfully yield the same result.  Also, the model was applied to
linear stability 
analyses of the magnetorotational instability in an accretion disk system, 
and was shown to reproduce the results in the previous study including
negative energy modes.

Although the Hamiltonian structure of the four-field reduced model was
not fully utilized in the linear stability analyses.  
However, the calculation of the perturbed energy was based on the
Hamiltonian formulation, and it was clearly shown that the collision of
positive and negative energy modes leads to instability.
The Hamiltonian structure should be useful if we consider more complex
problems such as a situation where   
a negative energy mode becomes unstable due to dissipation.

\begin{acknowledgments}
M.F. was supported by JSPS KAKENHI Grant Numbers JP21K03507 and JP24K06993. 
\end{acknowledgments}

\section*{Conflict of Interest}

The authors have no conflicts to disclose.

\section*{Author Contributions}

Masaru Furukawa:
Conceptualization (equal), 
Data Curation (lead),
Formal Analysis (equal),
Funding Acquisition (lead),
Investigation (equal),
Methodology (equal),
Project Administration (lead),
Resources (lead),
Software (lead),
Supervision (equal),
Validation (equal),
Visualization (lead),
Writing/Original Draft Preparation (lead),
Writing/Review \& Editing (equal)

Makoto Hirota:
Conceptualization (equal), 
Data Curation (supporting),
Formal Analysis (equal),
Funding Acquisition (supporting),
Investigation (equal),
Methodology (equal),
Project Administration (supporting),
Resources (supporting),
Software (supporting),
Supervision (equal),
Validation (equal),
Visualization (supporting),
Writing/Original Draft Preparation (supporting),
Writing/Review \& Editing (equal)

\section*{Data Availability Statement}

The data that support the findings of this study are available from the corresponding author upon reasonable request.

\appendix

\section{Calculation of mode energy}
\label{sec:energyCalculation}

This appendix describes how 
the perturbed energy of an eigenmode was calculated.
The calculation is based on Ref.~\onlinecite{Hirota-2008}.
For a reference, several expressions of the energy of perturbed modes
are compared in Ref.~\onlinecite{Fukumoto-2023}.
The perturbed energy calculation including Alfv\'en continua 
was formulated in a general manner in Ref.~\onlinecite{Hirota-2008-PoP}.

As described in Ref.~\onlinecite{Hirota-2008}, linearized equations of a
Hamiltonian system 
is written in the form  
\begin{equation}
 \frac{\pd \tilde{v}}{\pd t}
= ( \mathcal{A} \mathcal{H} + \mathcal{B} ) \tilde{v}.
\label{eq:linearizedEqGeneral}
\end{equation}
Here, 
\begin{align}
 \mathcal{H} 
&:=
 \left.
 \left( \frac{\delta^{2} H}{\delta v^{i} \delta v^{j}} \right)
 \right\vert_{v_{\mathrm{e}}},
\label{eq:calH-definition}
\\
 \mathcal{A} 
&:=
 \left. \mathcal{J} \right\vert_{v_{\mathrm{e}}},
\label{eq:calA-definition}
\\
 \mathcal{B}
&:=
 \left.
 \left(
   \left[
     \frac{\delta H}{\delta v^{k}} ,  
      \circ \, a^{ik}_{j}
   \right]
 \right)
 \right\vert_{v_{\mathrm{e}}},
\label{eq:calB-definition}
\end{align}
where $v_{\mathrm{e}}$ denotes an equilibrium.

For the specific model in this paper,
the perturbed variables are 
$v(r, \alpha ,t) := 
 ( U, u_{h}, \psi, B_{h}^{\star}
 )^{\mathrm{T}}$,
where tildes representing perturbation are omitted 
in this Appendix~\ref{sec:energyCalculation}
for simplicity.
The operators $\mathcal{A}$, $\mathcal{H}$,
and $\mathcal{B}$ are given by 
\begin{align}
 \mathcal{H} := & \,
\begin{pmatrix}
  -\mathcal{L}^{-1}
& 0
& 0
& 0
\\
  0
& g
& 0
& 0
\\
  0
& 0
& -\mathcal{L} - 2 M K f g
& -2 M K g
\\
  0
& 0
& -2 M K g
& 1 / g
\end{pmatrix},
%\label{eq:calH-definition}
\\
 \mathcal{A} := & \,
\begin{pmatrix}
  [ \circ , U + f u_{h} ] 
& [ \circ , u_{h} ] 
& [ \circ , \psi ] 
& [ \circ , g B_{h} ] 
\\
  [ \circ , u_{h} ] 
& 0
& 0
& [ \circ , \psi ] 
\\
  [ \circ , \psi ] 
& 0
& 0
& 0
\\
  [ \circ , g B_{h} ] 
& [ \circ , \psi ] 
& 0
& 0
\end{pmatrix},
\\
 \mathcal{B} := & \,
\begin{pmatrix}
  [ -\vphi , \circ ] 
& [ -\vphi , f \circ ] + [ g u_{h} , \circ ] 
& [ -\mathcal{L} \psi + f B_{h} , \circ ] + [ B_{h} , f \circ ] 
& [ B_{h} , \circ ] 
\\
  0
& [ -\vphi , \circ ] 
& [ B_{h} , \circ ] 
& 0
\\
  0
& 0
& [ -\vphi , \circ ] 
& 0
\\
  0
& 0
& [ -\vphi , f \circ ] + [ g u_{h} , \circ ] 
& [ -\vphi , \circ ] 
\end{pmatrix},
\end{align}
where equilibrium quantities are substituted in each element
of $\mathcal{H}$, $\mathcal{A}$ and $\mathcal{B}$.
The functions $f(r)$ and $g(r)$ are defined in 
Eqs.~(\ref{eq:f-definition-1}) and (\ref{eq:g-definition-1}),
respectively.  The operator $\mathcal{L}$ is defined 
in Eq.~(\ref{eq:calL-definition}).
For general definitions of the operators
$\mathcal{H}$, $\mathcal{A}$, and $\mathcal{B}$, see Ref.~\onlinecite{Hirota-2008}.

Let us introduce an adjoint equation \cite{Hirota-2008} as 
\begin{equation}
 \frac{\pd \zeta}{\pd t}
= ( \mathcal{H} \mathcal{A} - \mathcal{B}^{*} ) \zeta,
\label{eq:adjLinearizedEqGeneral}
\end{equation}
where an adjoint operators are defined through,
for example, 
\begin{equation}
 \left\langle
   \zeta, \mathcal{H} v
 \right\rangle
=
 \left\langle
   \mathcal{H}^{*} \zeta , v
 \right\rangle.
 \label{eq:adjcalH-definition}
\end{equation}
The relations 
$\mathcal{H}^{*} = \mathcal{H}$,
$\mathcal{A}^{*} = - \mathcal{A}$,
$(\mathcal{B}\mathcal{A})^{*} = \mathcal{B}\mathcal{A}$
were shown to hold, where the last one was derived from the Jacobi identity 
(\ref{eq:LiePoissonBracketJacobiIdentity}).
Using these properties, it was also shown that 
$ ( \mathcal{A} \mathcal{H} + \mathcal{B} ) \mathcal{A} =
 \mathcal{A} ( \mathcal{H} \mathcal{A} - \mathcal{B}^{*} )$.

Operating $\mathcal{A}$ on Eq.~(\ref{eq:adjLinearizedEqGeneral}),
we obtain 
\begin{equation}
 \frac{\pd ( \mathcal{A} \zeta ) }{\pd t}
= ( \mathcal{A} \mathcal{H} + \mathcal{B} ) ( \mathcal{A} \zeta )
\end{equation}
by using the relations of these operators.
Therefore, if we obtain a solution $\zeta$ 
of Eq.~(\ref{eq:adjLinearizedEqGeneral}),
$v = \mathcal{A} \zeta$ is a solution to 
Eq.~(\ref{eq:linearizedEqGeneral}).
This perturbation restricted on the range of $\mathcal{A}$ 
is called dynamically accessible.\cite{Morrison-1998}

Let us express the perturbed quantities by a Fourier series 
\begin{equation}
 v (r, \alpha ,t )
= 
 \sum_{\ell = -\infty}^{\infty} 
 v_{\ell}(r, t) \rme^{\rmi \ell K \alpha}.
\label{eq:vFourierSeriesExpantion}
\end{equation}
Substituting this expression into the linearized equation
(\ref{eq:linearizedEqGeneral}), 
we obtain 
\begin{align*}
 \frac{\pd}{\pd t}
 \sum_{\ell = -\infty}^{\infty} 
 v_{\ell}(r, t) \rme^{\rmi \ell K \alpha}
&=
 ( \mathcal{A} \mathcal{H} + \mathcal{B} ) 
 \sum_{\ell = -\infty}^{\infty} 
 v_{\ell}(r, t) \rme^{\rmi \ell K \alpha}
\\
&=
 -\rmi 
 \sum_{\ell = -\infty}^{\infty} 
 \mathcal{L}_{\ell}
 v_{\ell}(r, t) \rme^{\rmi \ell K \alpha},
\end{align*}
and thus
\begin{equation}
 \frac{\pd v_{\ell}(r, t)}{\pd t} 
= 
 - \rmi \mathcal{L}_{\ell} v_{\ell}(r, t),
 \label{eq:linearizedEqFourierMode}
\end{equation}
where the operator $\mathcal{L}_{\ell}$
is defined through
\begin{equation}
 ( \mathcal{A} \mathcal{H} + \mathcal{B} ) v
=:
 - \rmi \sum_{\ell = -\infty}^{\infty} \mathcal{L}_{\ell} 
   v_{\ell}(r, t) \rme^{\rmi \ell K \alpha}.
 \label{eq:calLl-definition} 
\end{equation}
Note that 
$\mathcal{L}_{\ell}$ is a real operator.
%, and
%it can be factored by $\ell$ originating from $\pd / \pd \alpha$ terms
%in the Poisson brackets in $\mathcal{A}$ and $\mathcal{B}$.

Similarly, a Fourier series expansion of $\zeta$,
\begin{equation}
 \zeta (r, \alpha ,t )
= 
 \sum_{\ell = -\infty}^{\infty} 
 \zeta_{\ell}(r, t) \rme^{\rmi \ell K \alpha}
\label{eq:zetaFourierSeriesExpantion}
\end{equation}
is substituted in the adjoint equation (\ref{eq:adjLinearizedEqGeneral}) 
to obtain 
\begin{equation}
 \frac{\pd \zeta_{\ell}(r, t)}{\pd t} 
= 
 - \rmi \mathcal{L}^{*}_{\ell} \zeta_{\ell}(r, t),
\label{eq:adjLinearizedEqFourierMode}
\end{equation}
where the operator $\mathcal{L}^{*}$ is defined through
\begin{equation}
 ( \mathcal{H} \mathcal{A} - \mathcal{B}^{*} ) \zeta
= 
 - \rmi \sum_{\ell = -\infty}^{\infty} \mathcal{L}^{*}_{\ell}
   \zeta_{\ell}(r, t) \rme^{\rmi \ell K \alpha}.
\label{eq:calLstl-definition} 
\end{equation}

Let us examine a relation between $v_{\ell}$ and $\zeta_{\ell}$.
An operator $\mathcal{A}_{\ell}$ is defined through
\begin{equation}
 \mathcal{A} \zeta(r, \alpha, t)
= 
 - \rmi \sum_{\ell = -\infty}^{\infty} \mathcal{A}_{\ell} 
   \zeta_{\ell}(r, t) \rme^{\rmi \ell K \alpha}.
\end{equation}
Comparing this expression with the Fourier series expansion of $v$
(\ref{eq:vFourierSeriesExpantion}), we obtain 
\begin{equation}
 v_{\ell} (r, t)
=
 - \rmi \mathcal{A}_{\ell} \zeta_{\ell} (r, t).
\label{eq:vlAlzetal}
\end{equation}

Now, by assuming the time dependence of $v_{\ell}(r, t)$
as $v_{\ell}(r) \rme^{-\rmi \omega_{\ell} t}$,
we obtain an eigenvalue problem
\begin{equation}
 \omega_{\ell} v_{\ell}(r) = \mathcal{L}_{\ell} v_{\ell}(r)
\label{eq:eigenvalueProblemFourierMode}
\end{equation}
from Eq.~(\ref{eq:linearizedEqFourierMode}).
Note that, if we found an eigenvalue $\omega_{\ell}$ and the corresponding
eigenfunction $v_{\ell}(r)$, the complex conjugate 
$\overline{\omega_{\ell}}$ is also an eigenvalue
and the corresponding eigenfunction is $\overline{v_{\ell}(r)}$.

Similarly, by assuming 
$\zeta_{\ell}(r, t) = \zeta_{\ell}(r) \rme^{-\rmi \omega_{\ell} t}$,
we obtain
\begin{equation}
 \omega_{\ell} \zeta_{\ell} = \mathcal{L}^{*}_{\ell} \zeta_{\ell}
\label{eq:adjEigenvalueProblemFourierMode}
\end{equation}
from the adjoint equation (\ref{eq:adjLinearizedEqFourierMode}).
If we found an eigenvalue $\omega_{\ell}$ and the corresponding
eigenfunction $\zeta_{\ell}(r)$, the complex conjugate 
$\overline{\omega_{\ell}}$ is also an eigenvalue
and the corresponding eigenfunction is $\overline{\zeta_{\ell}(r)}$.
Because of the relation (\ref{eq:vlAlzetal}),
we have a relation between eigenfunctions sharing a same $\omega$ as
$v_{\ell}(r) = -\rmi \mathcal{A}_{\ell} \zeta_{\ell}(r)$.
We will omit the argument $(r)$ of $v_{\ell}$ and $\zeta_{\ell}$
hereafter for simplicity of notation.

Suppose we found eigenvalues $\omega_{\ell n}$ and $\omega_{\ell m}$,
and the corresponding eigenfunctions are
$v_{\ell n}$ and $v_{\ell m}$
for Eq.~(\ref{eq:eigenvalueProblemFourierMode})
and 
$\zeta_{\ell n}$ and $\zeta_{\ell m}$
for Eq.~(\ref{eq:adjEigenvalueProblemFourierMode}),
where $n$ and $m$ are indices to distinguish the eigenvalues.
Then, integrating 
the eigenmode equation
$\omega_{\ell n} v_{\ell n} = \mathcal{L}_{\ell} v_{\ell n}$
multiplied by $\overline{\zeta_{\ell m}}$ inside the plasma volume, 
we obtain 
\begin{align*}
 \omega_{\ell n} 
 \left\langle
  \overline{\zeta_{\ell m}} , v_{\ell n} 
 \right\rangle
&= 
 \left\langle
 \overline{\zeta_{\ell m}} , \mathcal{L}_{\ell} v_{\ell n}
 \right\rangle
\nonumber
\\
&= 
 \left\langle
 \mathcal{L}^{*}_{\ell} \overline{\zeta_{\ell m}} , v_{\ell n}
 \right\rangle
\nonumber
\\
&= 
 \left\langle
  \overline{\omega_{\ell m}} \overline{\zeta_{\ell m}} , v_{\ell n}
 \right\rangle ,
\end{align*}
and thus
\begin{equation}
 ( \omega_{\ell n} - \overline{\omega_{\ell m}} )
 \left\langle
 \overline{\zeta_{\ell m}} , v_{\ell n}
 \right\rangle
 = 0.
\label{eq:orthogonality}
\end{equation}
Therefore, if 
$\overline{\omega^{\ell m}} \neq \omega_{\ell n}$,
the eigenfunction 
$\overline{\zeta^{\ell m}}$ 
is orthogonal to $v_{\ell n}$.
Its complex conjugate is also true.

Let us take an eigenmode with its eigenvalue 
$\omega_{\ell m}$ and the eigenfunction $v_{\ell n}$.
Then, the physical perturbation and its adjoint quantity should be
\begin{align}
 v(r, \alpha, t)
&=
 v_{\ell n} \rme^{-\rmi \omega_{\ell n}t} \rme^{\rmi \ell K \alpha}
+ \overline{v_{\ell n}} \rme^{\rmi \overline{\omega_{\ell n}} t}
 \rme^{-\rmi \ell K \alpha},
\label{eq:singleModeV}
\\
 \zeta(r, \alpha, t)
&=
 \zeta^{\ell n} \rme^{-\rmi \omega_{\ell n}t} \rme^{\rmi \ell K \alpha}
+ \overline{\zeta^{\ell n}} \rme^{\rmi \overline{\omega_{\ell n}} t} \rme^{-\rmi \ell K \alpha}
\nonumber
\\
&=
 \rmi \mathcal{A}_{\ell}^{-1}
 v_{\ell n} \rme^{-\rmi \omega_{\ell n}t} \rme^{\rmi \ell K \alpha}
- \rmi \mathcal{A}_{\ell}^{-1}
\overline{v_{\ell n}} \rme^{\rmi \overline{\omega_{\ell n}} t} \rme^{-\rmi \ell K \alpha}.
\end{align}
The perturbed mode energy is given by\cite{Hirota-2008}
\begin{align}
 \delta^{2} H
&=
 -\left\langle
 \frac{\pd v}{\pd t} , \zeta
  \right\rangle
\nonumber
\\
&=
 \omega_{\ell n} 
 \rme^{-\rmi ( \omega_{\ell n} - \overline{\omega_{\ell n}} ) t}
 \left\langle
   v_{\ell n} , \overline{\zeta_{\ell n}}
 \right\rangle
 + \text{c.c.},
\end{align}
where c.c. stands for complex conjugate.
This $\delta^{2} H$ is the difference of the energy $H$ in the
perturbed state  from that in the equilibrium, where the perturbation is
taken to be dynamically accessible.

If the eigenmode is a pure oscillation, $\omega_{\ell n}$ is real,
and $\omega_{\ell n} = \overline{\omega_{\ell n}}$.
Therefore, $\delta^{2} H$ reduces
\begin{equation}
\delta^{2} H
=
 2 \, \Re \,
 \omega_{\ell n}
 \left\langle
   v_{\ell n} , \overline{\zeta_{\ell n}}
 \right\rangle.
\label{eq:modeEnergyIntegral}
\end{equation}
If the eigenmode is unstable, 
$\omega_{\ell n} \neq \overline{\omega_{\ell n}}$, and therefore
$\delta^{2} H = 0$ by the orthogonality (\ref{eq:orthogonality}).
%If $\delta^{2} H \neq 0$ for an unstable mode, the energy of the
%perturbation grows exponentially in time even though the system
%conserves the energy.
Since the quadratic form $\delta^{2} H$ is a constant of motion for the
linearized Hamiltonian system, it must be zero if an exponentially
growing eigenmode is substituted (otherwise $\delta^{2} H$ would grow in
time). 
In any case, the mode energy can be calculated by 
Eq.~(\ref{eq:modeEnergyIntegral}).

Lastly, $\zeta_{\ell n}$ is obtained by using 
\begin{equation}
 \mathcal{A}_{\ell}^{-1}
=
 \frac{r}{\ell \psi^{\pr}}
 \begin{pmatrix}
   0 & 0 & 1 & 0
\\
   0 & 0 &
   -\frac{\left( g B_{h} \right)^{\pr}}{\psi^{\pr}}  
   & 1
\\
   1 
   & 
   -\frac{\left( g B_{h} \right)^{\pr}}{\psi^{\pr}}  
   & 
   -\frac{ ( U + f u_{h} )^{\pr} \psi^{\pr} - 2 u_{h}^{\pr} ( g B_{h} )^{\pr} }
         { (\psi^{\pr})^{2} }
   &
   - \frac{u_{h}^{\pr}}{\psi^{\pr}}
\\
  0 & 1 & - \frac{u_{h}^{\pr}}{\psi^{\pr}} & 0
\end{pmatrix},
\label{eq:calAlinv}
\end{equation}
where equilibrium quantities are substituted.
Note that the determinant of $\mathcal{A}_{\ell}$
is $( \ell \psi^{\pr} / r )^{4}$.
If the determinant does not vanish, we do not need to solve the adjoint
eigenvalue problem 
Eq.~(\ref{eq:adjEigenvalueProblemFourierMode}) to obtain 
$\zeta_{\ell n}$, 
but we only need to solve the eigenvalue problem 
Eq.~(\ref{eq:eigenvalueProblemFourierMode}).

Lastly, $\delta^{2} H$ should be normalized appropriately since the
eigenfunction can be scaled arbitrarily.
%A numerical solution to the the eigenvalue problem 
%Eq.~(\ref{eq:eigenvalueProblemFourierMode}) is output with an arbitrary
%normalization of the eigenfunction. 
%Therefore, it is appropriate to normalize $\delta^{2} H$ to eliminate
%the arbitrariness. 
The results presented in Sec.~\ref{subsec:negativeEnergymode}
used the following normalization:
\begin{equation}
 \tilde{H}
:=
 \frac{\delta^{2} H}
 {  \frac{1}{2}
 \int \td V \,
 \frac{1}{K_{0}^{2}r^{2}}
 \left(
    | \nab \tilde{\vphi} |^{2}
  + | \tilde{u}_{h} |^{2}
  + | \nab \tilde{\psi} |^{2}
  + | \tilde{B}_{h} |^{2}
 \right)  }.
\label{eq:normalizedModeEnergy}
\end{equation}

%aipnum4-2.bst 2019-01-14 (MD) hand-edited version of apsrev4-1.bst

\end{document}